\documentclass[journal]{IEEEtran}

\newtheorem{theorem}{Theorem}
\newtheorem{proposition}{Proposition}

\usepackage{booktabs}
\usepackage{bm}
\usepackage{graphicx}
\usepackage[cmex10]{amsmath}
\usepackage{cite}
\usepackage{amssymb}
\usepackage{subfigure}
\usepackage{extarrows}
\usepackage[letterspace = -2]{microtype}
\usepackage{algorithm}
\usepackage{algpseudocode}
\usepackage{algpascal}
\usepackage{float}

\usepackage{color}
\usepackage{xcolor}

\newfloat{algorithm}{t}{lop}

\begin{document}

\title{Large-Scale Multi-Antenna Multi-Sine \\Wireless Power Transfer
}
\author{
Yang~Huang
and Bruno~Clerckx,~\IEEEmembership{Senior Member,~IEEE}
\thanks{A preliminary version has been presented in the IEEE International Workshop on Signal Processing Advances in Wireless Communications 2015 \cite{HC16SPAWC}.}
\thanks{Y. Huang and B. Clerckx are with the Department of Electrical and Electronic Engineering, Imperial College London, London SW7 2AZ, United Kingdom (e-mail: \{y.huang13, b.clerckx\}@imperial.ac.uk). The work of Y. Huang was supported by CSC Imperial Scholarship. This work has been partially supported by the EPSRC of UK, under grant EP/P003885/1.}
}

\maketitle

\vspace{-0.2cm}
\begin{abstract}
Wireless Power Transfer (WPT) is expected to be a technology reshaping the landscape of low-power applications such as the Internet of Things, RF identification (RFID) networks, etc. To that end, multi-antenna multi-sine waveforms adaptive to the Channel State Information (CSI) have been shown to be a promising building block of WPT. However, the current design is computationally too complex to be applied to large-scale WPT, where the transmit signal is sent across a large number (tens) of antennas and frequencies.
In this paper, we derive efficient single-user and multi-user algorithms based on a generalizable optimization framework, in order to design transmit waveforms that maximize the weighted-sum/minimum rectenna DC output voltage. The study highlights the significant effect of the nonlinearity introduced by the rectification process on the design of waveforms in single/multi-user systems.
Interestingly, in the single-user case, the optimal spatial domain beamforming, obtained prior to the frequency domain power allocation optimization, turns out to be Maximum Ratio Transmission (MRT).
On the contrary, in the general multi-user weighted sum criterion maximization problem, the spatial domain beamforming optimization and the frequency domain power allocation optimization are coupled. Assuming channel hardening, low-complexity algorithms are proposed based on asymptotic analysis, to maximize the two criteria. The structure of the asymptotically optimal spatial domain precoder can be found prior to the optimization. The performance of the proposed algorithms is evaluated. Numerical results confirm the inefficiency of the linear model-based design for the single and multi-user scenarios.
It is also shown that as nonlinear model-based designs, the proposed algorithms can benefit from an increasing number of sinewaves at a computational cost much lower than the existing method. Simulation results highlight the significant benefits of the large-scale WPT architecture to boost the end-to-end power transfer efficiency and the transmission range.
\end{abstract}


\section{Introduction}
Low-power applications such as the Internet of Things and radio frequency identification networks are expected to benefit from far-field Wireless Power Transfer (WPT) systems\cite{CM16}. WPT utilizes a dedicated source to radiate electromagnetic energy through a wireless channel and a rectenna at the receiver to convert this energy into DC power.

In the study of far-field WPT, though circuit-level designs have been intensively studied in an effort to improve the RF-to-DC conversion efficiency of a rectifier, the RF literature has revealed that this efficiency is also a function of the input waveforms and can be boosted by multi-sine WPT i.e. transmitting a superposition of sinewaves over frequencies with uniform frequency spacing\cite{BBFCGC15, CM16, BCCG13, BC11}. In the context of microwave theory, methods for improving the energy transfer efficiency have been studied in smart power beaming \cite{CM16} (not leveraging multi-sine signals) and spatially combined multi-sine signals\cite{BBFCGC15}. The main limitation of those methods is not only the lack of a systematic approach to design waveforms, but also the fact that they operate without Channel State Information (CSI) at the Transmitter (CSIT)\footnote{Attempts have been made to obtain CSI for WPT, e.g. \cite{ZZ15, XZ14, ZCZ16}.}, which in communications is usually used to turn multipath fading into a benefit for users.

One challenge in designing waveform strategies adaptive to CSIT is constructing an analytical rectenna model.
Although most off-the-shelf rectifier models in the context of microwave theory are able to provide insights into the accurate rectification process, non-closed forms or highly complex structures in these models\cite{LW15, VMD15} make it hard to derive efficient algorithms for WPT.
By contrast, \cite{ZZH13} built a simplified (but inaccurate \cite{ABDV05,ZCZ16, CBmar16TSParxiv}) model by truncating the Taylor expansion of the Shockley diode equation to the 2nd-order term, where the rectenna DC output current is linearly proportional to the average incident RF power. This model is referred to as a \emph{linear} model.
Nevertheless, the linear model may not properly describe the rectenna behaviour.
In the scenario where multiple frequency components can be utilized for WPT, the linear model favours a waveform strategy where the transmit power is allocated to a single frequency component \cite{ZCZ16, CBmar16TSParxiv}. Unfortunately, this contradicts experimental results that show the benefits of a power allocation over multiple frequency components to boost the RF-to-DC conversion efficiency \cite{BBFCGC15, CM16, BCCG13, BC11}.
The reason is that the 4th-order truncation is necessary for characterizing the basic diode rectification process\cite{BC11}. To balance the accuracy and complexity of the optimization, \cite{CBmar16TSParxiv} truncates the Taylor expansion of the diode equation to the $n_o$\,th-order term ($n_o \geq 4$), yielding a \emph{nonlinear} rectenna model.
Based on this nonlinear model, \cite{CBmar16TSParxiv} developed waveform optimization methods for multi-antenna multi-sine WPT.
Circuit simulations in \cite{CBmar16TSParxiv} validate the newly developed nonlinear model, highlight the inaccuracy of the linear model and confirm the superiority of optimized waveforms over various baselines. It is shown that the linear model favours narrowband transmission, while the nonlinear model favours a power allocation over multiple frequencies. The nonlinear model leads to a completely different spectrum usage compared to the linear model\cite{ZCZ16, CBmar16TSParxiv}.

Interestingly, leveraging the nonlinear model, the scaling law for multi-sine WPT in \cite{CBmar16TSParxiv} shows that the DC output current scales linearly with the number of sinewaves.
This sheds interests in a large-scale design of WPT with many sinewaves and transmit antennas, so as to jointly benefit from the rectifier non-linearity, channel frequency selectivity and a beamforming gain.
The use of a large number of antennas is somehow reminiscent of massive MIMO in communications. Unfortunately, the nonlinear model-based waveform optimizations in \cite{CBmar16TSParxiv} turn out to be posynomial maximization problems, which are solved by reversed Geometric Programming (GP).
These reversed GP algorithms, though applicable to any truncation order $n_o$, suffer exponential complexity\cite{MChiang05} and hence cannot be applied to large-scale designs. Moreover, the reversed GP approach relies on optimizing magnitudes for a given set of phases. In the multi-user setup, phases and amplitudes have to be jointly optimized, which limits the applicability of the reversed GP approach \cite{CBmar16TSParxiv}.

This paper studies the large-scale WPT, where the name stems from the fact that the number of transmit antennas and sinewaves can be large, e.g. tens of transmit antennas and tens of sinewaves. To the authors' best knowledge, this is the first paper looking at efficient nonlinear model-based waveform optimizations for a large-scale WPT architecture. The paper studies not only low-complexity waveform optimization algorithms but also the effect of the nonlinearity introduced by the rectification process on waveform designs. The main contributions are listed as follows.

First, we derive a novel analytical nonlinear rectenna model, by applying the approach in \cite{Wetenkamp83}. The new model characterizes the rectenna DC output voltage $v_\text{out}$. Interestingly, multiplying this $v_\text{out}$ model by a constant leads to the DC current model in \cite{CBmar16TSParxiv}. They are equivalent in terms of optimization, even though they rely on different physical assumptions.

Second, we develop a computationally efficient optimization framework to address waveform designs involving the Taylor series-based 4th-order truncation model. This optimization framework is also general enough to cope with single-user and multi-user setups. Although \cite{CBmar16TSParxiv} opens interesting perspectives for WPT where waveforms are adaptive to CSI, the RF rectenna model (based on a Taylor series expansion truncated to the $n_o$th order) and thereby the formulated optimizations lead to the highly complex reversed GP. Moreover, the reversed GP approach cannot guarantee the optimality of the waveform design in the multi-user scenario. To avoid the exponential complexity of the reversed GP and therefore come up with a computationally tractable method for waveform optimizations but also be able to derive optimal multi-user waveforms, we use a Taylor series-based 4th order truncation model and convert the proposed RF rectenna model into a compact expression: a real-valued function of complex vector variables. However, the expression is essentially a quartic function and results in intractable waveform optimizations. These problems are NP-hard in general, while off-the-shelf algorithms for standard quartic problems are inapplicable or cannot guarantee to converge to stationary points\cite{LNQY09, LZ10, BLQX12}. As a systematic treatment, the developed optimization framework introduces auxiliary variables and exploits convex relaxations \cite{LZ10}, such that the quartic objective can be reduced to a nonconvex quadratic constraint in an equivalent problem. Then, the nonconvex constraint is linearized, and the equivalent problem is iteratively approximated. Following this, a variety of convex optimization techniques (e.g. Successive Convex Approximation (SCA)\cite{MW78}, rank reduction\cite{HP10, HP14}, etc.) can be used to solve the approximate problem. The waveform optimization framework is derived for a single-user WPT and is then generalized to multi-user WPT systems. The proposed approach is shown to provide a much lower complexity compared to the reversed GP in \cite{CBmar16TSParxiv} for any number of antennas/sinewaves. Nevertheless, it becomes very significant as the number of antennas/sinewaves grows large.

Third, assuming perfect CSIT, we propose waveform optimization algorithms to maximize a general multi-user weighted sum $v_\text{out}$. We reveal that in the particular single-user case, the optimal spatial domain beamforming turns out to be Maximum Ratio Transmission (MRT). Hence, only the power allocation across frequencies needs to be optimized by an algorithm. The comparison of this single-user waveform design and the linear model-based design again confirms the inefficiency of the linear model \cite{CBmar16TSParxiv}.
On the contrary, in the general multi-user scenario, we show that the power allocation across frequencies affects the optimal spatial domain beamforming, due to the nonlinear rectification.
Hence, an algorithm is proposed to jointly optimize both the spatial domain beamforming and the power allocation across frequencies. In the presence of more antennas and sinewaves, a channel hardening-exploiting algorithm is proposed, assuming that channel hardening happens among the spatial domain channel vectors across users and frequencies. It is revealed that the asymptotically optimal spatial domain precoder is a linear combination of the users' spatial domain channels. Hence, the algorithms optimize power allocation across frequencies and users. As a nonlinear model-based design, the frequency domain power optimization can benefit from an increasing number of sinewaves. It is also shown that increasing the number of antennas/sinewaves can boost the power transfer efficiency and enlarge the transmission range.

Fourth, for the sake of fairness among users, we propose waveform optimization algorithms to maximize the minimum $v_\text{out}$. In the presence of a large number of antennas and sinewaves, we also propose channel hardening-exploiting algorithms. We show that the minimum $v_\text{out}$ achieved by these fairness-aware algorithms can be improved by increasing the number of sinewaves.

\emph{Organization}: Section \ref{SecSystemModel} discusses the system model. Section \ref{SecSU_WO} optimizes waveform for a single user. Sections \ref{SecWaveOptAlgos} and \ref{SecMaxMinAlgos} respectively maximize the weighted-sum/minimum $v_\text{out}$. Section \ref{SecSimResults} evaluates the performance. Conclusions are drawn in Section \ref{SecConclu}.

\emph{Notations}: Matrices and vectors are in bold capital and bold lower cases, respectively. The notations $(\cdot)^T$, $(\cdot)^\star$, $(\cdot)^\ast$, $(\cdot)^H$, $\text{Tr}\{\cdot\}$, $\|\cdot\|$, $\|\cdot\|_F$ and $|\cdot|$ represent the transpose, optimized solution, conjugate, conjugate transpose, trace, 2-norm, Frobenius norm and absolute value, respectively. The notation $\mathbf{A} \succeq 0$ means that $\mathbf{A}$ is positive-semidefinite.

\section{System Model}\label{SecSystemModel}
\subsection{Signal Transmission through Wireless Channels}\label{SecSignalTrans}
In the studied WPT system, an $M$-antenna base station (BS) serves $K$ single-antenna users (or rectennas) by delivering multi-sine deterministic power signals over $N$ frequencies. It is assumed that all spatial/frequency domain channel gains remain constant during the transmission, and the BS has perfect CSIT. The complex scalar channel gain between the $m$\,th antenna and the user $q$ at the $n$\,th frequency is designated as $h_{q,(n-1)M + m}$ (for $q = 1,\ldots,K$, $n = 1, \ldots, N$ and $m = 1, \ldots, M$), which is collected into $\mathbf{h}_q \in \mathbb{C}^{MN \times 1}$. Hence, $\mathbf{h}_q = [\mathbf{h}_{q,1}^T, \ldots,$ $\mathbf{h}_{q,N}^T]^T$, where $\mathbf{h}_{q,n} = [h_{q,(n-1)M+1}, \ldots, h_{q,(n-1)M+M}]^T$ describes the spatial domain channel gains at the $n$\,th frequency.

By collecting the magnitude and the phase of the RF complex signal at angular frequency $\omega_n$ into a complex variable $s_{(n-1)M+m}$, the complex version of the signal transmitted by the $m$\,th BS antenna can be formulated as $\tilde{x}_m(t) \triangleq \sum_{n = 1}^N s_{(n-1)M+m} e^{j\omega_n t}$. Hence, the RF signal transmitted by antenna $m$ is $x_m(t) = \sqrt{2} \text{Re}\{\tilde{x}_m(t)\}$. The variable $s_{(n-1)M+m}$ is collected into a waveform precoder $\mathbf{s} \in \mathbb{C}^{MN \times 1}$, such that $\mathbf{s} = [\mathbf{s}_1^T, \ldots, \mathbf{s}_N^T]^T$ and $\mathbf{s}_n = [s_{(n-1)M+1}, \ldots, s_{(n-1)M+M}]^T$ which characterizes the signals transmitted at angular frequency $\omega_n$. Note that $\omega_n \!=\! \omega_1 \! + \! (n\!-\!1)\Delta_\omega$, for $n \! = \! 1, \ldots, N$ and $\omega_1 \!> \! (N \! - \! 1)\Delta_\omega/2$. Suppose that the BS transmit power is constrained by $P$, such that $\|\mathbf{s}\|^2 = \sum_{n = 1}^N \sum_{m = 1}^M |s_{(n-1)M + m}|^2 \leq P$. The complex RF signal through the channel between the $m$\,th transmit antenna and the $q$\,th user can be written as
\begin{equation}
\label{EqYq_tilde}
\tilde{y}_{q, m}(t) = \textstyle{\sum_{n = 1}^N} s_{(n - 1)M+m} h_{q,(n-1)M+m} e^{j\omega_n t}\,.
\end{equation}
Hence, the RF signal input into the antenna at user $q$ is
\begin{equation}
\label{EqYq}
y_q(t) = \sqrt{2} \text{Re}\left\{ \tilde{y}_q(t) \right\} = \sqrt{2} \text{Re}\left\{\textstyle{\sum_{m = 1}^M} \tilde{y}_{q,m}(t)\right\}\,.
\end{equation}

\subsection{The Nonlinear Rectifying Process}\label{SecNonlinearRectfPro}
\begin{figure}[!t]
\centering
\includegraphics[width = 1.8in]{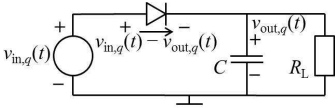}
\caption{Rectifier with a single-series diode, where $R_{\text{L}}$ is the load.}
\label{FigEHcirAna}
\end{figure}
Similarly to \cite{CBmar16TSParxiv}, we assume a lossless antenna and perfect matching. Given an antenna impedance $R_{\text{ant}} = 50\Omega$, the rectifier input voltage can be obtained as $v_{\text{in},q}(t) = y_q(t)\sqrt{R_{\text{ant}}}$.

Accounting for a rectifier as shown in Fig. \ref{FigEHcirAna} and applying the method in the RF literature \cite{Wetenkamp83}, we now construct a novel nonlinear model, which characterizes the rectifier output voltage $v_{\text{out},q}(t)$. This method assumes that the rectifier output current $i_{\text{dc},q} \approx 0$, due to the low-power input and the high impedance load \cite{Wetenkamp83, BBFCGC15, BCCG13, BC11}. On the other hand, the method of \cite{CBmar16TSParxiv} assumes an operating voltage drop at the diode. Despite the above difference, the achieved nonlinear models are equivalent in terms of optimization.

The $v_{\text{out},q}(t)$ model is derived by manipulating the Shockley equation, which describes the instantaneous diode current
\begin{equation}
\label{EqIdq}
i_{\text{d},q}(t) = i_\text{s}\left[\exp\left(\frac{v_{\text{in},q}(t) - v_{\text{out},q}(t)}{n_\text{i} V_\text{T}}\right) -  1\right]\,,
\end{equation}
where $i_\text{s}$, $V_\text{T}$ and $n_\text{i}$ represent the saturation current, the thermal voltage and the ideality factor (set to 1 for simplicity), respectively. We assume that the capacitor $C$ functions as an ideal low-pass filter $f_\text{LPF}(\cdot)$, which removes non-DC components in $i_{\text{d},q}(t)$ and the rectenna output voltage $v_{\text{out},q}(t)$. Hence, $v_{\text{out},q}(t)$ reduces to a DC voltage $v_{\text{out},q}$, and
\begin{equation}
\label{EqIdcq}
i_{\text{dc},q} = f_\text{LPF}(i_{\text{d},q}(t)) = i_\text{s} e^{-\frac{v_{\text{out},q}}{n_\text{i} V_\text{T}}} \cdot f_\text{LPF}\left( e^{\frac{v_{\text{in},q}(t)}{n_\text{i} V_\text{T}}}\right) -  i_\text{s}\,.
\end{equation}
Recall that $i_{\text{dc},q} \approx 0$ \cite{Wetenkamp83, BBFCGC15, BCCG13, BC11}. Hence, $v_{\text{out},q}$ can be written as a function of $v_{\text{in},q}(t)$. Eq. (\ref{EqIdcq}) highlights that the nonlinear behavior of the diode comes from the term $\exp(\frac{v_{\text{in},q}(t)}{n_\text{i} V_\text{T}})$. In order to obtain a tractable analytical model, this term is addressed by the small-signal analysis method: we apply Taylor expansion to this term (where $\frac{v_{\text{in},q}(t)}{n_\text{i} V_\text{T}}$ is regarded as a whole) at 0 quiescent point. As it is revealed in the RF literature \cite{BC11} and the waveform design \cite{CBmar16TSParxiv} that the 4th-order truncation is necessary for characterizing the basic rectification, we truncate the Taylor series to the 4th-order term. Therefore,
\begin{equation}
\label{EqVoutq_Scalar1}
v_{\text{out},q} = n_\text{i} V_\text{T} \ln \left[1 + \frac{f_\text{LPF}\left(v_{\text{in},q}^2(t)\right)}{2 n_\text{i}^2 V_\text{T}^2} + \frac{f_\text{LPF}\left(v_{\text{in},q}^4(t)\right)}{24 n_\text{i}^4 V_\text{T}^4}\right]\,,
\end{equation}
where the odd order terms $f_\text{LPF}(v_{\text{in},q}(t))/(n_\text{i} V_\text{T})$ and $f_\text{LPF}(v_{\text{in},q}^3(t))/(6 n_\text{i}^3 V_\text{T}^3)$ vanish and are removed. By exploiting $\ln(1 + x) \simeq x$ and combining (\ref{EqVoutq_Scalar1}) and $v_{\text{in},q}(t)$, we achieve $v_{\text{out},q}$ as a function of $y_q(t)$
\begin{equation}
\label{EqVoutq_Scalar2}
v_{\text{out},q} = \beta_2 f_\text{LPF}\left(y_q^2(t)\right) + \beta_4 f_\text{LPF}\left(y_q^4(t)\right)\,,
\end{equation}
where $\beta_2 = R_\text{ant}/(2 n_\text{i} V_\text{T})$ and $\beta_4 = R_\text{ant}^2/(24 n_\text{i}^3 V_\text{T}^3)$. Interestingly, multiplying the above $v_{\text{out},q}$ by $i_\text{s}/(n_\text{i} V_\text{T})$ achieves nothing else than the model $z_{DC}$ (truncated to the 4th-order term) in \cite{CBmar16TSParxiv}. Hence, the two models are equivalent in terms of optimization. Due to this and the fact that the model in \cite{CBmar16TSParxiv} has been validated by simulations in various rectifier configurations\cite{CBmar16TSParxiv, CB17}, we do not conduct circuit simulations in this paper. The above model is based on small signal analysis and valid only for a diode operating in the nonlinear region. If $v_{\text{in},q}(t)$ is so large that the diode series resistance dominates the diode behaviour and the diode $I$-$V$ characteristic is linear\cite{BCCG13}, the assumptions made for approximation as well as the Taylor series-based model does not hold.

Then, $v_{\text{out},q}$ can be expressed as a function of the transmit waveform, by combining (\ref{EqYq_tilde}), (\ref{EqYq}) and (\ref{EqVoutq_Scalar2}). In the term $f_{\text{LPF}}\left(y_q^2(t)\right) = f_{\text{LPF}}\big(\text{Re}\{ \tilde{y}_q(t)\tilde{y}_q(t) \! + \tilde{y}_q(t)\tilde{y}_q^\ast(t) \}\big)$, where $\text{Re}\{\tilde{y}_q(t)\tilde{y}_q(t)\}$ can be removed as it only contains the non-DC components. Hence,
\begin{IEEEeqnarray}{l}
\label{EqFlpf_yq2}
f_{\text{LPF}} \left(y_q^2(t)\right) {}={} \sum_{n = 1}^N \sum_{m_1, m_2} \left[s_{(n - 1)M + m_1} h_{q,(n - 1)M + m_1} \cdot\right. \nonumber\\
\quad\left. s^\ast_{(n - 1)M + m_2} h^\ast_{q, (n - 1)M + m_2}\right] = \textstyle{\sum_{n = 1}^N} \mathbf{s}_n^H \mathbf{h}_{q,n}^\ast \mathbf{h}_{q,n}^T \mathbf{s}_n,
\end{IEEEeqnarray}
where $m_1, m_2 \in \{1, \ldots, M\}$. Similarly,
\begin{IEEEeqnarray}{l}
\label{EqFlpf_yq4}
f_\text{LPF}\!\left(y_q^4(t)\right) = \frac{3}{2} f_\text{LPF}\left( \text{Re}\left\{ \tilde{y}_q(t) \tilde{y}_q(t) \tilde{y}_q^\ast(t) \tilde{y}_q^\ast(t) \right\} \right) \IEEEyesnumber\IEEEyessubnumber\\
= \frac{3}{2} \!\sum_{\substack{{n_1, n_2, n_3, n_4}\\{n_1 \!-\!  n_3\!  =\!  -\!  (n_2\!  -\!  n_4)}}}\! \mathbf{s}_{n_3}^H \mathbf{h}_{q, n_3}^\ast  \mathbf{h}_{q, n_1}^T \mathbf{s}_{n_1} \! \cdot \! \mathbf{s}_{n_4}^H  \mathbf{h}_{q,n_4}^\ast \mathbf{h}_{q,n_2}^T \mathbf{s}_{n_2},\quad \IEEEyessubnumber \label{EqFlpf_yq4_2}
\end{IEEEeqnarray}
where $n_1, n_2, n_3, n_4\! \in\! \{1, \dots, N\}$ and $m_1, m_2, m_3, m_4 \! \in \! \{1, \ldots, M\}$. Substituting (\ref{EqFlpf_yq2}) and (\ref{EqFlpf_yq4}) into (\ref{EqVoutq_Scalar2}) yields
\begingroup\makeatletter\def\f@size{9}\check@mathfonts
\def\maketag@@@#1{\hbox{\m@th\small\normalfont#1}}%
\begin{IEEEeqnarray}{l}
\label{Eqv_outInhomo}
v_{\text{out},q} = \beta_2 \textstyle{\sum_{n = 1}^N} \mathbf{s}_n^H \mathbf{h}_{q,n}^\ast \mathbf{h}_{q,n}^T \mathbf{s}_n + \nonumber\\
\frac{3}{2} \beta_4 \! \sum_{\substack{{n_1, n_2, n_3, n_4}\\{n_1 \!-\!  n_3\!  =\!  -\!  (n_2\!  -\!  n_4)}}} \! \mathbf{s}_{n_3}^H \! \mathbf{h}_{q,n_3}^\ast \! \mathbf{h}_{q,n_1}^T \! \mathbf{s}_{n_1} \!\cdot\! \mathbf{s}_{n_4}^H \! \mathbf{h}_{q,n_4}^\ast \! \mathbf{h}_{q,n_2}^T \! \mathbf{s}_{n_2}.
\end{IEEEeqnarray}
\endgroup
\begin{figure}[!t]
\centering
\includegraphics[width = 3.0in]{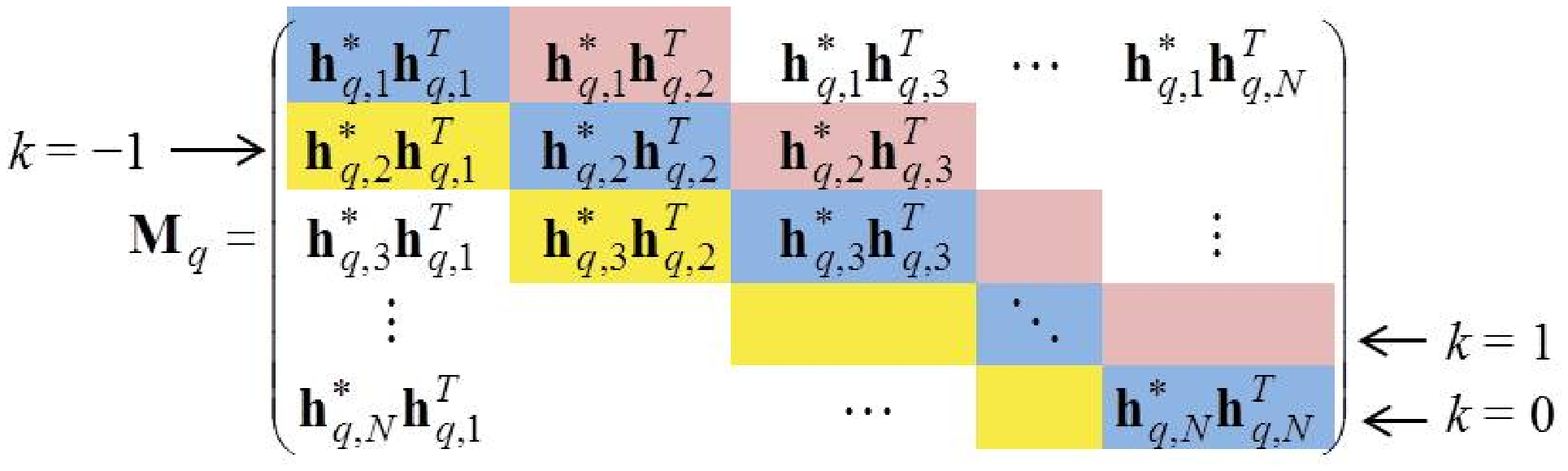}
\caption{$\mathbf{M}_{q,1}$ is the above matrix only maintaining the block diagonal (whose index is $k=1$) in pink, while all the other blocks are set as $\mathbf{0}_{M\times M}$.}
\label{FigM_mat}
\end{figure}
\emph{A compact expression:} The above (\ref{Eqv_outInhomo}) can be transformed into a more compact form, by introducing $MN$-by-$MN$ matrices $\mathbf{M}_q \! \triangleq \! \mathbf{h}_q^\ast \mathbf{h}_q^T$ and $\mathbf{M}_{q,k}$. As shown in Fig. \ref{FigM_mat}, $k \! \in \! \{1, \ldots, N\!-\!1\}$ is the index of the $k$\,th block diagonal above the main block diagonal (whose index $k=0$) of $\mathbf{M}_q$, while $k \! \in \! \{-\!(N\!-\!1), \ldots, -1\}$ is the index of the $|k|$\,th block diagonal below the main block diagonal. Given a certain $k$, $\mathbf{M}_{q,k}$ is generated by retaining the $k$\,th block diagonal of $\mathbf{M}_q$ but setting all the other blocks as $\mathbf{0}_{M\times M}$. For $k\neq 0$, the non-Hermitian matrix $\mathbf{M}_{q,-k}\! =\! \mathbf{M}_{q,k}^H$, while $\mathbf{M}_{q,0}\! \succeq\! 0$. Thus, $v_{\text{out},q}$ can be written as
\begingroup\makeatletter\def\f@size{9}\check@mathfonts
\def\maketag@@@#1{\hbox{\m@th\small\normalfont#1}}%
\begin{IEEEeqnarray}{rcl}
\label{EqFuncVoutq}
v_{\text{out},q} & {}={} & \beta_2 \mathbf{s}^H \mathbf{M}_{q,0} \mathbf{s} \! + \! \frac{3}{2}\beta_4 \mathbf{s}^H \mathbf{M}_{q,0} \mathbf{s}\!\left(\mathbf{s}^H \mathbf{M}_{q,0}\mathbf{s}\right)^H \! + \nonumber\\
&&\! 3\beta_4 \textstyle{\sum_{k = 1}^{N-1}} \mathbf{s}^H \mathbf{M}_{q,k} \mathbf{s} \!\left(\mathbf{s}^H \mathbf{M}_{q,k} \mathbf{s}\right)^H\,.
\end{IEEEeqnarray}
\endgroup
This compact form cannot be generalized to a higher-order truncation model. For a higher-order truncation model, waveform optimizations have to be solved by reversed GP\cite{CBmar16TSParxiv}. Nevertheless, the key benefit of this compact form is the potential to solve the waveform design problem in a computationally much more efficient way than the GP approach and to solve the optimal multi-user waveforms.

\section{Single-User Waveform Optimization}\label{SecSU_WO}
\subsection{A Frequency Domain Power Allocation Problem}\label{SecAFreqDomPwrAlloc}
To maximize the DC output voltage shown in (\ref{Eqv_outInhomo}) or (\ref{EqFuncVoutq}) for a single-user (i.e. $K=1$ and $q=1$) WPT, the waveform optimization problem can be formulated as
\begingroup\makeatletter\def\f@size{9}\check@mathfonts
\def\maketag@@@#1{\hbox{\m@th\small\normalfont#1}}%
\begin{equation}
\label{ProbOriginalSUwpt}
\max_{\mathbf{s}} \left\{ v_{\text{out},1}: \|\mathbf{s}\|^2 \leq P \right\}\,.
\end{equation}
\endgroup

As defined in Section \ref{SecSignalTrans}, $\mathbf{s}$ collects the spatial domain precoder across frequencies. Intuitively, finding an optimal $\mathbf{s}^\star$ for problem (\ref{ProbOriginalSUwpt}) amounts to optimizing both the normalized spatial domain beamforming and the power allocation across frequencies. The following theorem reveals that the optimal spatial domain beamforming for a single-user WPT is essentially MRT. This means that only the power allocation across frequencies would be optimized by an iterative algorithm.
\begin{theorem}
\label{TheoSUwptMRT}
In problem (\ref{ProbOriginalSUwpt}), the optimal spatial domain precoder $\mathbf{s}_n^\star$ at frequency $n$ has a structure of $\mathbf{s}_n^\star \triangleq \xi_n \mathbf{\tilde{s}}_n$, where the optimal (normalized) spatial domain beamforming $\mathbf{\tilde{s}}_n = e^{j\phi_n}  \mathbf{h}_{1,n}^\ast/\|\mathbf{h}_{1,n}\|$ and $\xi_n$ is generally a complex weight.
\end{theorem}
\begin{IEEEproof}
Please refer to Appendix \ref{AppTheoSUwptMRT} for details.
\end{IEEEproof}

\emph{Remark:} Appendix \ref{AppTheoSUwptMRT} reveals that to maximize $v_{\text{out},1}$, though the optimal $\xi_n^\star$ and $\phi_n$ can be real and zero respectively as in \cite{CBmar16TSParxiv}, non-zero $\phi_n$ and complex $\xi_n^\star$ (meanwhile, $\xi_n^\star e^{j\phi_n}$ is complex) can be found to achieve the same $v_{\text{out},1}$, e.g. for $N = 3$, $\phi_n$ and the phase of $\xi_n^\star$ (which is designated as $\angle \xi_n^\star$) only need to satisfy $2(\phi_2 + \angle \xi_2^\star)= \phi_1 + \angle \xi_1^\star + \phi_3 + \angle \xi_3^\star$. Remarkably, Theorem 1 implies that in the presence of multi-sine WPT, phase modulated information signals could be embedded into energy signals, while the $v_\text{out}$ achieved by these signals can be equal to that achieved by pure WPT.

In the following, without loss of generality, we redefine $\mathbf{s}_n^\star \triangleq \xi_n \cdot \mathbf{h}_{1,n}^\ast/ \|\mathbf{h}_{1,n}\|$. Note that $|\xi_n|^2$ is the power allocated to the precoder $\mathbf{s}_n$ at frequency $n$. Substituting the optimal structure of $\mathbf{s}_n$ into (\ref{Eqv_outInhomo}) yields
\begingroup\makeatletter\def\f@size{9}\check@mathfonts
\def\maketag@@@#1{\hbox{\m@th\small\normalfont#1}}%
\begin{IEEEeqnarray}{rcl}
\label{Eqv_outInhomoSU}
v_{\text{out},1} &{}={}& \beta_2 \sum_{n = 1}^N |\xi_n|^2 \|\mathbf{h}_{1,n}\|^2 + \frac{3}{2} \beta_4 \! \sum_{\substack{{n_1, n_2, n_3, n_4}\\{n_1\!-\!  n_3\!  =\!  - \! (\!n_2\!  - \! n_4\!)}}} \! \xi_{n_3}^\ast \|  \mathbf{h}_{1,n_3} \| \cdot \nonumber\\
&&\| \mathbf{h}_{1,n1} \| \xi_{n_1} \xi_{n_4}^\ast  \| \mathbf{h}_{1,n_4}\| \|\mathbf{h}_{1,n_2}\| \xi_{n_2}\,.
\end{IEEEeqnarray}
\endgroup

Before formulating the optimization problem with respect to $\xi_n$, we formulate (\ref{Eqv_outInhomoSU}) in a more compact form similar to that in Section \ref{SecNonlinearRectfPro}. Hence, we introduce $N$-by-$N$ matrices $\mathbf{M}^{\prime\prime}_k$ and $\mathbf{M}^{\prime\prime} \triangleq \mathbf{h}_\text{norm}\mathbf{h}_\text{norm}^T$, where $\mathbf{h}_\text{norm} = [\|\mathbf{h}_{1,1}\|, \ldots, \|\mathbf{h}_{1,N}\|]^T$. Given $k$, $\mathbf{M}^{\prime\prime}_k$ is generated by retaining the $k$\,th diagonal\footnote{The subscript $k \! \in \! \{1, \ldots, N\!-\!1\}$ in $\mathbf{M}^{\prime\prime}_k$ is the index of the $k$\,th diagonal above the main diagonal (whose index $k=0$) of $\mathbf{M}^{\prime\prime}$, while $k \! \in \! \{-\!(N\!-\!1), \ldots, -1\}$ is the index of the $|k|$\,th diagonal below the main diagonal.} of $\mathbf{M}^{\prime\prime}$ but setting all the other entries as 0. Defining $\mathbf{p} \triangleq [\xi_1, \ldots, \xi_N]^T$, (\ref{Eqv_outInhomoSU}) can be written as
\begingroup\makeatletter\def\f@size{9}\check@mathfonts
\def\maketag@@@#1{\hbox{\m@th\small\normalfont#1}}%
\begin{IEEEeqnarray}{rcl}
\label{Eqv_outHomoSU}
v_{\text{out},1} &{}={}& \beta_2 \mathbf{p}^H\mathbf{M}_0^{\prime\prime}\mathbf{p} + \frac{3}{2}\beta_4\mathbf{p}^H\mathbf{M}_0^{\prime\prime}\mathbf{p}\left(\mathbf{p}^H\mathbf{M}_0^{\prime\prime}\mathbf{p}\right)^H +\nonumber\\
&& 3\beta_4 \textstyle{\sum_{k = 1}^{N-1}} \mathbf{p}^H \mathbf{M}_k^{\prime\prime} \mathbf{p} \left(\mathbf{p}^H \mathbf{M}_k^{\prime\prime} \mathbf{p}\right)^H\,.
\end{IEEEeqnarray}
\endgroup
Hence, according to Theorem \ref{TheoSUwptMRT}, problem (\ref{ProbOriginalSUwpt}) boils down to a frequency domain optimization problem, whose epigraph form can be formulated as
\begingroup\makeatletter\def\f@size{9}\check@mathfonts
\def\maketag@@@#1{\hbox{\m@th\small\normalfont#1}}%
\begin{IEEEeqnarray}{cl}
\label{Prob_QuarticProb}
\min_{\gamma_0,\,  \mathbf{p}} \quad & \gamma_0 \IEEEyesnumber\IEEEyessubnumber\\
\text{s.t.} \,& - v_{\text{out},1}\left(\mathbf{p}\right)  - \gamma_0 \leq 0 \,, \IEEEyessubnumber \label{EqHighDegPolyConstSU}\\
& \mathbf{p}^H\mathbf{p} \leq P\,. \IEEEyessubnumber
\end{IEEEeqnarray}
\endgroup
The function $v_{\text{out},1}$ is a quartic polynomial, which in general makes problem (\ref{Prob_QuarticProb}) NP-hard \cite{LNQY09, LZ10, BLQX12}. To address the quartic polynomial, auxiliary variables $t_{k}$ (for $k = 0, \ldots, N\!-\!1$) are introduced, such that $\mathbf{p}^H \mathbf{M}_k^{\prime\prime} \mathbf{p} = t_k$, and the quartic constraint (\ref{EqHighDegPolyConstSU}) can be reduced to a quadratic constraint. However, for $k \neq 0$, $\mathbf{M}_k^{\prime\prime}$ is not hermitian, and the term $\mathbf{p}^H \mathbf{M}_k^{\prime\prime} \mathbf{p}$ is essentially a bilinear function, which may also lead to a NP-hard problem\cite{VBS99}. To handle this, a rank-1 matrix variable $\mathbf{X}$ is introduced to linearize the bilinear term, such that $\mathbf{p}^H \mathbf{M}_k^{\prime\prime} \mathbf{p} = \text{Tr}\{\mathbf{M}_k^{\prime\prime} \mathbf{p} \mathbf{p}^H\} = \text{Tr}\{\mathbf{M}_k^{\prime\prime} \mathbf{X}\} = t_k$. Then, defining $\mathbf{t} = [t_0,\ldots,t_{N-1}]^T$,
\begingroup\makeatletter\def\f@size{9}\check@mathfonts
\def\maketag@@@#1{\hbox{\m@th\small\normalfont#1}}%
\begin{equation}
\label{EqMatA0}
\mathbf{A}_0 =  diag\{-3\beta_4/2, -3\beta_4, \ldots, -3\beta_4\}  \preceq  0
\end{equation}
\endgroup
and $g(\mathbf{t}) \triangleq \mathbf{t}^H \mathbf{A}_0 \mathbf{t} = - 3 \beta_4 t_0 t_0^\ast/2 - 3\beta_4 \textstyle{\sum_{k=1}^{N-1}} t_k t_k^\ast$, problem (\ref{Prob_QuarticProb}) can be equivalently reformulated as
\begingroup\makeatletter\def\f@size{9}\check@mathfonts
\def\maketag@@@#1{\hbox{\m@th\small\normalfont#1}}%
\begin{IEEEeqnarray}{cl}
\label{EquiProb_MaxVoutSU}
\min_{\gamma_0,\, \mathbf{t},\, \mathbf{X}\succeq 0} \, & \gamma_0 \IEEEyesnumber\IEEEyessubnumber\\
\text{s.t.} \,& - \beta_2 t_0 + g(\mathbf{t}) \! - \! \gamma_0 \! \leq \! 0, \IEEEyessubnumber \label{Eq_MaxVoutSU_RankConstRelaxed_nonCVXquadrConst}\\
& \text{Tr}\{\mathbf{M}_k^{\prime\prime} \mathbf{X}\} = t_k\,,\quad \forall k\,, \IEEEyessubnumber \label{EqConst_SU_t_k}\\
& \text{Tr}\{\left[\mathbf{M}_k^{\prime\prime}\right]^H\mathbf{X}\} = t_k^\ast\,,\quad \forall k\neq0\,, \IEEEyessubnumber \label{EqConstConjugatet_SU_k}\\
& \text{Tr}\{\mathbf{X}\} \leq P\,, \IEEEyessubnumber \label{EqTxPwrConst_SU}\\
& \text{rank}\{\mathbf{X}\} = 1\,.\IEEEyessubnumber \label{EqEpiMaxVoutSU_RankConst}
\end{IEEEeqnarray}
\endgroup

\subsection{Single-User Waveform Optimization Algorithm}\label{SecWeighSumSCAwaveformOpt}
Unfortunately, the nonconvex quadratic constraint (\ref{Eq_MaxVoutSU_RankConstRelaxed_nonCVXquadrConst}) and the rank constraint (\ref{EqEpiMaxVoutSU_RankConst}) make problem (\ref{EquiProb_MaxVoutSU}) NP-hard in general. We first relax the rank constraint, solving
\begingroup\makeatletter\def\f@size{9}\check@mathfonts
\def\maketag@@@#1{\hbox{\m@th\small\normalfont#1}}%
\begin{IEEEeqnarray}{cl}
\label{Epi_Problem_MaxVoutSU_RankConstRelaxed}
\min_{\gamma_0,\, \mathbf{t},\, \mathbf{X}\succeq 0} \quad & \left\{ \gamma_0: \text{(\ref{Eq_MaxVoutSU_RankConstRelaxed_nonCVXquadrConst}), (\ref{EqConst_SU_t_k}), (\ref{EqConstConjugatet_SU_k}), and (\ref{EqTxPwrConst_SU})} \right\}\,.
\end{IEEEeqnarray}
\endgroup
By this means, solving problem (\ref{EquiProb_MaxVoutSU}) amounts to finding a rank-constrained $\mathbf{X}^\star$ of the relaxed problem (\ref{Epi_Problem_MaxVoutSU_RankConstRelaxed}).

\subsubsection{Solving the Relaxed Problem}\label{SecWeighSumSCAbasedAlgSU}
As problem (\ref{Epi_Problem_MaxVoutSU_RankConstRelaxed}) suffers from the nonconvex quadratic constraint (\ref{Eq_MaxVoutSU_RankConstRelaxed_nonCVXquadrConst}), the solution of (\ref{Epi_Problem_MaxVoutSU_RankConstRelaxed}) is approximated iteratively by SCA. At iteration $l$, the nonconvex term $g(\mathbf{t})$ in (\ref{Eq_MaxVoutSU_RankConstRelaxed_nonCVXquadrConst}) is approximated at $\mathbf{t}^{(l - 1)}$, which is the $\mathbf{t}^\star$ optimized at iteration $(l - 1)$, as a linear function by its first-order Taylor expansion\cite{AL10}
\begingroup\makeatletter\def\f@size{9}\check@mathfonts
\def\maketag@@@#1{\hbox{\m@th\small\normalfont#1}}%
\begin{equation}
\label{Eq_g_tilde}
\tilde{g}\left(\mathbf{t}, \mathbf{t}^{(l - 1)}\right) \triangleq 2\text{Re}\left\{\left[\mathbf{t}^{(l - 1)}\right]^H \mathbf{A}_0 \mathbf{t}\right\} - \left[\mathbf{t}^{(l - 1)}\right]^H \mathbf{A}_0 \mathbf{t}^{(l - 1)}\,.
\end{equation}
\endgroup
Then, the $l$\,th approximate problem (AP) can be written as
\begingroup\makeatletter\def\f@size{9}\check@mathfonts
\def\maketag@@@#1{\hbox{\m@th\small\normalfont#1}}%
\begin{IEEEeqnarray}{cl}
\label{ApproxConvProblem_MaxVoutSU}
\min_{\gamma_0, \mathbf{t}, \mathbf{X}\succeq 0} \, & \gamma_0 \IEEEyesnumber\IEEEyessubnumber \label{EqObjApproxCvxProbSU}\\
\text{s.t.} \,& - \beta_2 t_0 + \tilde{g}\left(\mathbf{t}, \mathbf{t}^{(l - 1)}\right) -  \gamma_0  \leq  0, \IEEEyessubnumber \label{EqApproxCvxProbSU_LinearizedConst}\\
& \text{(\ref{EqConst_SU_t_k}), (\ref{EqConstConjugatet_SU_k}), and (\ref{EqTxPwrConst_SU})}\,,\nonumber
\end{IEEEeqnarray}
\endgroup
which is a Semidefinite Problem (SDP).
Substituting (\ref{EqConst_SU_t_k}) and (\ref{EqConstConjugatet_SU_k}) into (\ref{EqApproxCvxProbSU_LinearizedConst}) shows that problem (\ref{ApproxConvProblem_MaxVoutSU}) is essentially equivalent to
\begingroup\makeatletter\def\f@size{9}\check@mathfonts
\def\maketag@@@#1{\hbox{\m@th\small\normalfont#1}}%
\begin{IEEEeqnarray}{cl}
\label{ApproxConvProblem_MaxVoutSU_Equiv}
\min_{\mathbf{X}\succeq 0} \, & \left\{\text{Tr}\{\mathbf{A}_1^{\prime\prime}\mathbf{X}\}: \text{Tr}\{\mathbf{X}\} \leq P \right\}\,,
\end{IEEEeqnarray}
\endgroup
where $\mathbf{A}_1^{\prime\prime} \triangleq \mathbf{C}_1^{\prime\prime} + [\mathbf{C}_1^{\prime\prime}]^H$ is Hermitian, and
\begingroup\makeatletter\def\f@size{9}\check@mathfonts
\def\maketag@@@#1{\hbox{\m@th\small\normalfont#1}}%
\begin{equation}
\label{Eq_C1doublePrime}
\mathbf{C}_1^{\prime\prime} = -\frac{\beta_2 + 3 \beta_4 t^{(l - 1)}_0}{2} \mathbf{M}_0^{\prime\prime} - 3\beta_4 \textstyle{\sum_{k=1}^{N\!-\!1}} \left[t^{(l - 1)}_k \right]^\ast \mathbf{M}_k^{\prime\prime}\,.
\end{equation}
\endgroup
Applying \cite[Proposition 3.5]{HP10} shows that problem (\ref{ApproxConvProblem_MaxVoutSU_Equiv}) has, among others, a rank-1 optimal solution $\mathbf{X}^\star$. Due to the equivalence, this $\mathbf{X}^\star$ also satisfies the Karush-Kuhn-Tucker (KKT) conditions of (\ref{ApproxConvProblem_MaxVoutSU}). As (\ref{ApproxConvProblem_MaxVoutSU}) is convex, this $\mathbf{X}^\star$ is also a rank-1 global optimum of the AP (\ref{ApproxConvProblem_MaxVoutSU}).

\subsubsection{Achieving a Rank-1 $\mathbf{X}^\star$ of the AP (\ref{ApproxConvProblem_MaxVoutSU})}\label{SecSolveApproxProbSU}
We can solve (\ref{ApproxConvProblem_MaxVoutSU_Equiv}) by the interior-point method and obtain a high-rank $\mathbf{X}$, from which a rank-1 $\mathbf{X}^\star$ can be derived by rank reduction\cite[Algorithm 1]{HP10}. However, solving the SDP (\ref{ApproxConvProblem_MaxVoutSU_Equiv}) causes complexity of $O(1)(2+2N)^{1/2}N^2\big(5N^4 + 8N^3 + N^2 +1\big)$\cite{BN01}. Fortunately, the following method yields a closed-form rank-1 $\mathbf{X}^\star$ of (\ref{ApproxConvProblem_MaxVoutSU}), with reduced complexity.
\begin{proposition}
\label{PropRank1soluSU}
Problem (\ref{ApproxConvProblem_MaxVoutSU_Equiv}) can yield, among others, a rank-1 optimal solution $\mathbf{X}^\star = \mathbf{x}^\star[\mathbf{x}^\star]^H$, where $\mathbf{x}^\star = \sqrt{P}[\mathbf{U}_{\mathbf{A}_1^{\prime\prime}}]_{\text{min}}$ and $[\mathbf{U}_{\mathbf{A}_1^{\prime\prime}}]_{\text{min}}$ is the eigenvector corresponding to the minimum eigenvalue of $\mathbf{A}_1^{\prime\prime}$.
\end{proposition}
\begin{IEEEproof}
See Appendix \ref{AppPropRank1soluSU} for details.
\end{IEEEproof}
In Proposition \ref{PropRank1soluSU}, performing Eigenvalue Decomposition (EVD) for $\mathbf{A}_1^{\prime\prime}$ by the QR algorithm yields complexity of
$O\big(N^3\big)$\cite{Parlett00}. The overall algorithm is summarized in Algorithm \ref{AlgSCA_SU}.
\begin{algorithm}
\caption{Single-User (SU) WPT Algorithm}\label{AlgSCA_SU}
\begin{algorithmic}[1]
\State \textbf{Initialize} Set $l = 0$, and generate $\mathbf{X}^{(0)}$, $\mathbf{t}^{(0)}$ and $\gamma_0^{(0)}$;
\Repeat
    \State $l = l + 1$;
    \State Compute $\mathbf{A}_1^{\prime\prime}$; $\mathbf{x}^\star = \sqrt{P}\left[\mathbf{U}_{\mathbf{A}_1^{\prime\prime}}\right]_{\text{min}}$; $\mathbf{X}^\star = \mathbf{x}^\star[\mathbf{x}^\star]^H$;
    \State Update $\mathbf{X}^{(l)} = \mathbf{X}^\star$; Update $t_k^{(l)}$ $\forall k$ by (\ref{EqConst_SU_t_k});
\Until{\|\mathbf{X}^{(l)} - \mathbf{X}^{(l-1)}\|_F/\|\mathbf{X}^{(l)}\|_F \leq \epsilon}
\State $\mathbf{p}^\star = \mathbf{x}^\star$ and $\mathbf{s}_n^\star = [\mathbf{p}^\star]_{n,1} \cdot \mathbf{h}_{1,n}^\ast/ \|\mathbf{h}_{1,n}\|$ for $n = 1,\ldots, N$.
\end{algorithmic}
\end{algorithm}
\begin{theorem}
\label{TheoConvergenceAlgSCA_SU}
If the eigenvectors of a given $\mathbf{A}_1^{\prime\prime}$ are uniquely attained, Algorithm \ref{AlgSCA_SU} converges to a stationary point of problem (\ref{EquiProb_MaxVoutSU}).
\end{theorem}
\begin{IEEEproof}
As shown in \cite[Theorem 1]{MW78}, the convergence of SCA is proved under the assumption that the minimizer converges to a limit point. However, theoretically, the eigenvectors of a given Hermitian matrix $\mathbf{A}_1^{\prime\prime}$ are not unique, and $\mathbf{X}^{(l)}$ in Algorithm \ref{AlgSCA_SU} may not converge to a limit point. Therefore, similarly to \cite{BH03} and \cite{HC16twc}, we establish the convergence under the condition that the eigenvectors of a given $\mathbf{A}_1^{\prime\prime}$ are uniquely attained. See Appendix \ref{AppTheoConvergenceAlgSCA_SU} for the detailed proof.
\end{IEEEproof}

Though \cite[Algorithm 2]{CBmar16TSParxiv} also converges to a stationary point, it suffers exponential complexity. By contrast, Algorithm \ref{AlgSCA_SU} features low per-iteration complexity, i.e. $O\big(N^3\big)$, and inherits low convergence time property from the SCA for nonconvex Quadratically Constrained Quadratic Problems (QCQPs)\cite{MHGKS15}. In Section \ref{SecSimResults_PwrConst}, Table \ref{Tab_SUWPT_vs_ReversedGP} draws a comparison of the elapsed running time for SU WPT and reversed GP\cite{CBmar16TSParxiv}, demonstrating the high computational efficiency of SU WPT in practice.

\section{Weighted Sum $v_\text{out}$ Maximization Algorithms}\label{SecWaveOptAlgos}
\subsection{Weighted Sum $v_\text{out}$ Maximization}\label{SecGeneWirelessChan}
This subsection addresses the waveform optimization problem for a $K$-user system, aiming at maximizing the weighted sum output voltage. Defining $w_q \geq 0$ as the weight for user $q$, the problem can be formulated as
\begingroup\makeatletter\def\f@size{9}\check@mathfonts
\def\maketag@@@#1{\hbox{\m@th\small\normalfont#1}}%
\begin{IEEEeqnarray}{cl}
\label{Prob_MaxVoutWsum}
\max_{\mathbf{s}} \quad &\left\{ \textstyle{\sum_{q=1}^K} w_q\cdot v_{\text{out},q}: \|\mathbf{s}\|^2 \leq P\right\} \label{EqObj_Prob_MaxVoutWsum}
\end{IEEEeqnarray}
\endgroup

\emph{Remark :} In contrast to the single-user problem (\ref{ProbOriginalSUwpt}), it is hard to find the optimal spatial domain beamforming for the $K$-user problem (\ref{Prob_MaxVoutWsum}) prior to optimizing the frequency domain power allocation, due to the nonlinear rectification process. The intuition is revealed by a toy example as follows. The objective function of (\ref{EqObj_Prob_MaxVoutWsum}) can be expressed as
\begingroup\makeatletter\def\f@size{9}\check@mathfonts
\def\maketag@@@#1{\hbox{\m@th\small\normalfont#1}}%
\begin{IEEEeqnarray}{l}
\label{EqObj_Prob_MaxVoutWsum_analy}
\sum_{q=1}^K w_q\cdot v_{\text{out},q} = \beta_2  \sum_{n = 1}^N  \mathbf{s}_n^H \left[\sum_{q = 1}^K w_q \mathbf{h}_{q,n}^\ast \mathbf{h}_{q,n}^T \right] \mathbf{s}_n + \frac{3}{2} \beta_4 \cdot \nonumber\\
\sum_{\substack{{n_1, n_2, n_3, n_4}\\{n_1 \!-\! n_3\! =\! - \! (n_2\! -\! n_4)}}}\! \mathbf{s}_{n_3}^H \! \left[\! \sum_{q \!= 1}^K \! w_q \! \mathbf{h}_{q,\!n_3}^\ast \! \mathbf{h}_{q,\!n_1}^T \! \mathbf{s}_{n_1} \! \mathbf{s}_{n_4}^H \! \mathbf{h}_{q,n_4}^\ast \! \mathbf{h}_{q,n_2}^T \!\right]\! \mathbf{s}_{n_2}.
\end{IEEEeqnarray}
\endgroup
As shown in (\ref{EqObj_Prob_MaxVoutWsum_analy}), the 4th-order term (which is multiplied by $\beta_4$) is a sum of polynomials. The term is so complicated that we could not gain intuition from it. Therefore, we take the case of $N = 3$ as a toy example. Then, the polynomial in the 4th-order term with respect to $n_2 = n_3 = 1$ can be written as
\begin{equation}
\label{EqToyEg}
f_{\beta_4}\! =\! \mathbf{s}_1^H\! \left[\textstyle{\sum_{q=1}^K} w_q \mathbf{h}_{q,1}^\ast \! \left[ \textstyle{\sum_{n  =1}^3} \mathbf{h}_{q,n}^T \mathbf{s}_n \mathbf{s}_n^H \mathbf{h}_{q,n}^\ast \right]\! \mathbf{h}_{q,1}^T \right]\! \mathbf{s}_1.
\end{equation}
Define $\mathbf{s}_n \triangleq \xi_n \mathbf{\tilde{s}}_n $, where $\mathbf{\tilde{s}}_n$ represents the spatial domain beamforming, and $\xi_n$ is a complex weight related to the frequency domain power allocation. Given $\mathbf{\tilde{s}}_2$ and $\mathbf{\tilde{s}}_3$, if we would like to find the optimal $\mathbf{\tilde{s}}_1^\star$ that maximizes (\ref{EqToyEg}), the problem can be formulated as
\begin{equation}
\label{ProbToyEg}
\arg\max_{\mathbf{\tilde{s}}_1} \mathbf{s}_1^H \! \left[\textstyle{\sum_{q=1}^K} w_q \mathbf{h}_{q,1}^\ast \!\left[ \textstyle{\sum_{n=1}^3} |\xi_n|^2 |\theta_{q,n}|^2 \! \right] \! \mathbf{h}_{q,1}^T\right] \! \mathbf{s}_1\,,
\end{equation}
where $\theta_{q,n} \triangleq \mathbf{h}_{q,n}^T \mathbf{\tilde{s}}_n$. It can be seen that in the single-user case i.e. $K = 1$, the optimal $\mathbf{\tilde{s}}_1^\star$ is MRT. However, when $K >1$, as $|\xi_n|^2$ is coupled with $|\theta_{q,n}|^2$, the frequency domain power allocation affects the optimal structure of $\mathbf{\tilde{s}}_1$.

Therefore, we directly optimize $\mathbf{s}$ for problem (\ref{Prob_MaxVoutWsum}). The quartic $v_{\text{out},q}\,\forall q$ are tackled by introducing auxiliary variables in the same way as we transform (\ref{Prob_QuarticProb}) into (\ref{EquiProb_MaxVoutSU}). Defining $\mathbf{t}_q \triangleq [t_{q,0},\ldots,t_{q,N-1}]^T$  and $g_q(\mathbf{t}_q) \triangleq \mathbf{t}_q^H \mathbf{A}_0 \mathbf{t}_q$ (where $\mathbf{A}_0$ is defined in (\ref{EqMatA0})), problem (\ref{Prob_MaxVoutWsum}) can be equivalently recast as
\begin{IEEEeqnarray}{cl}
\label{Epi_Problem_MaxVout}
\min_{\gamma_1, \{\mathbf{t}_q\}_{q=1}^K, \mathbf{X}\succeq 0} \, & \gamma_1 \IEEEyesnumber\IEEEyessubnumber\\
\text{s.t.} & \textstyle{\sum_{q\!=\!1}^K} w_q \! \left(- \! \beta_2 t_{q,0}\! +\! g_q(\mathbf{t}_q)\right) \! - \! \gamma_1 \! \leq \! 0, \IEEEyessubnumber \label{Eq_MaxVout_RankConstRelaxed_nonCVXquadrConst}\\
& \text{Tr}\{\mathbf{M}_{q,k} \mathbf{X}\} = t_{q,k}\,,\quad \forall q, k\,, \IEEEyessubnumber \label{EqConst_t_qk}\\
& \text{Tr}\{\mathbf{M}_{q,k}^H \mathbf{X}\} = t_{q,k}^\ast\,,\quad \forall q, k\neq0\,, \IEEEyessubnumber \label{EqConstConjugatet_qk}\\
& \text{Tr}\{\mathbf{X}\} \leq P\,, \IEEEyessubnumber \label{EqTxPwrConst}\\
& \text{rank}\{\mathbf{X}\} = 1\,.\IEEEyessubnumber \label{EqEpiMaxVoutRankConst}
\end{IEEEeqnarray}
In order to solve (\ref{Epi_Problem_MaxVout}), we apply the approach used in Section \ref{SecWeighSumSCAbasedAlgSU}. That is, we relax the rank constraint and then solve the relaxed problem by SCA. Specifically, at iteration $l$, $g_q(\mathbf{t}_q)$ in (\ref{Eq_MaxVout_RankConstRelaxed_nonCVXquadrConst}) can be approximated (or linearized) as
\begin{equation}
\label{Eq_tilde_g_q}
\tilde{g}_q\!\left(\mathbf{t}_q, \mathbf{t}_q^{(l\!-\!1)}\!\right)\! \triangleq \! 2\text{Re}\!\left\{\left[\mathbf{t}_q^{(l\! -\! 1)}\right]^H\! \mathbf{A}_0 \mathbf{t}_q\!\right\}\! -\! \left[\mathbf{t}_q^{(l\! -\! 1)}\right]^H\! \mathbf{A}_0 \mathbf{t}_q^{(l \!-\! 1)}\,.
\end{equation}
Note that (\ref{Eq_tilde_g_q}) has the property $\tilde{g}_q(\mathbf{t}_q^{(l)}\!, \mathbf{t}_q^{(l)})\! =\! g_q(\mathbf{t}_q^{(l)}) \!\leq\! \tilde{g}_q(\mathbf{t}_q^{(l)}\!, \mathbf{t}_q^{(l-1)})$. Then, the $l$\,th convex AP can be obtained as
\begin{IEEEeqnarray}{cl}
\label{ApproxConvProblem_MaxVout}
\min_{\gamma_1\!,\{\!\mathbf{t}_q\!\}_{q\!=\!1}^K,\mathbf{X}\!\succeq 0} & \gamma_1 \IEEEyesnumber\IEEEyessubnumber \label{EqObjApproxCvxProb}\\
\text{s.t.} & \textstyle{\sum_{q\! =\! 1}^K} w_q \! \left(\! -\! \beta_2 t_{q,0} \! + \! \tilde{g}_q\!\left(\mathbf{t}_q, \mathbf{t}_q^{(l \! - \! 1)}\right)\!\right)\! \leq \! \gamma_1 \,,\quad \IEEEyessubnumber \label{EqApproxCvxProb_LinearizedConst}\\
& \text{(\ref{EqConst_t_qk}), (\ref{EqConstConjugatet_qk}), and (\ref{EqTxPwrConst})}\,,\nonumber
\end{IEEEeqnarray}
which is equivalent to
\begin{IEEEeqnarray}{cl}
\label{ApproxConvProblem_MaxVoutEquiv}
\min_{\mathbf{X}\succeq 0} \, & \left\{\text{Tr}\{\mathbf{A}_1\mathbf{X}\}: \text{Tr}\{\mathbf{X}\} \leq P \right\}\,,
\end{IEEEeqnarray}
where {\small$\mathbf{A}_1 \triangleq \mathbf{C}_1 + \mathbf{C}_1^H$} and {\small$\mathbf{C}_1 \! = \! \sum_{q = 1}^K \! w_q \!\big(\! -\frac{\beta_2 \! + \! 3\beta_4t^{(l\!-\!1)}_{q,0}}{2}\mathbf{M}_{q,0} \! - \! 3\beta_4 \! \sum_{k=1}^{N\!-\!1} \![t^{(l \!- \!1)}_{q,k}]^\ast \!\mathbf{M}_{q,k} \! \big)$}. According to Proposition \ref{PropRank1soluSU}, problem (\ref{ApproxConvProblem_MaxVoutEquiv}) can yield a rank-1 solution. The algorithm is summarized in Algorithm \ref{AlgSCA}, where performing EVD for $\mathbf{A}_1$ can have complexity of $O((MN)^3)$.
Similarly to Theorem \ref{TheoConvergenceAlgSCA_SU}, it can be shown that if the eigenvectors of a given $\mathbf{A}_1$ are uniquely attained, Algorithm \ref{AlgSCA} converges to a stationary point of problem (\ref{Epi_Problem_MaxVout}). However, \cite[Algorithm 4]{CBmar16TSParxiv} suffers exponential complexity but could not guarantee optimality.
\begin{algorithm}
\caption{Weighted Sum (WSum) Algorithm}\label{AlgSCA}
\begin{algorithmic}[1]
\State \textbf{Initialize} Set $l = 0$, and generate $\mathbf{X}^{(0)}$, $\{\mathbf{t}_q^{(0)}\}_{q = 1}^K$;
\Repeat
    \State $l = l + 1$;
    \State Compute $\mathbf{A}_1$; $\mathbf{x}^\star = \sqrt{P}\left[\mathbf{U}_{\mathbf{A}_1}\right]_{\text{min}}$; $\mathbf{X}^\star = \mathbf{x}^\star[\mathbf{x}^\star]^H$; \label{AlgSCALine4}
    \State Update $\mathbf{X}^{(l)} = \mathbf{X}^\star$; Update $t_{q,k}^{(l)}$ $\forall q,k$ by (\ref{EqConst_t_qk});
\Until{\|\mathbf{X}^{(l)} - \mathbf{X}^{(l-1)}\|_F/\|\mathbf{X}^{(l)}\|_F \leq \epsilon}
\State $\mathbf{s}^\star = \mathbf{x}^\star$. \label{AlgSCALine7}
\end{algorithmic}
\end{algorithm}

\subsection{Simplified Weighted Sum $v_\text{out}$ Maximization}\label{SecSimpWSum}
This subsection proposes a simplified $K$-user weighted sum $v_\text{out}$ algorithm, where the spatial domain beamforming is obtained in a closed form, based on a linear model. Specifically, the linear model with respect to the $K$-user weighted sum $v_\text{out}$ can be formulated as the 2nd-order truncation of (\ref{EqObj_Prob_MaxVoutWsum_analy}), i.e. the term in (\ref{EqObj_Prob_MaxVoutWsum_analy}) multiplied by $\beta_2$. Considering this linear model, in order to maximize the weighted sum criterion, the optimal spatial domain beamforming $\mathbf{w}_n$ (for $\|\mathbf{w}_n\| = 1$) is shown to be the dominant eigenvector of $\sum_{q = 1}^K w_q \mathbf{h}_{q,n}^\ast \mathbf{h}_{q,n}^T$. Then, considering such a \emph{linear model}-based spatial domain beamforming design, the frequency domain power allocation is optimized based on the \emph{nonlinear} model.

Given $\mathbf{w}_n$, the precoder at frequency $n$ is $\mathbf{s}_n = \xi_n\mathbf{w}_n$, for $\xi_n \in \mathbb{C}$. Similarly to (\ref{Eqv_outHomoSU}), we write $v_{\text{out},q}$ as a function of $\xi_n$ in a compact form, by introducing matrices {\small$\mathbf{M}_q^{\prime\prime\prime}$} and {\small$\mathbf{M}_{q,k}^{\prime\prime\prime}$}. It is defined that {\small$\mathbf{M}_q^{\prime\prime\prime} \triangleq \mathbf{h}_{\text{e},q} \mathbf{h}_{\text{e},q}^H$}, where {\small$\mathbf{h}_{\text{e},q} \triangleq [\mathbf{w}_1^H \mathbf{h}_{q,1}^\ast, \ldots, \mathbf{w}_N^H \mathbf{h}_{q,N}^\ast]^T$}. Given $k$, $\mathbf{M}_{q,k}^{\prime\prime\prime}$ is generated by retaining the $k$\,th diagonal \footnote{The subscript $k \! \in \! \{1, \ldots, N\!-\!1\}$ in $\mathbf{M}_{q,k}^{\prime\prime\prime}$ is the index of the $k$\,th diagonal above the main diagonal (whose index $k=0$) of $\mathbf{M}_q^{\prime\prime\prime}$, while $k \! \in \! \{-\!(N\!-\!1), \ldots, -1\}$ is the index of the $|k|$\,th diagonal below the main diagonal.} of $\mathbf{M}_q^{\prime\prime\prime}$ but setting all the other entries as 0. Therefore, by defining $\mathbf{p} \triangleq [\xi_1, \ldots, \xi_N]^T$,
{\small$v_{\text{out},q} = \beta_2 \mathbf{p}^H \mathbf{M}_{q,0}^{\prime\prime\prime} \mathbf{p} + \frac{3}{2} \beta_4 \mathbf{p}^H \mathbf{M}_{q,0}^{\prime\prime\prime} \mathbf{p} ( \mathbf{p}^H \mathbf{M}_{q,0}^{\prime\prime\prime} \mathbf{p} )^H + 3\beta_4 \textstyle{\sum_{k = 1}^{N-1}} \mathbf{p}^H \cdot \mathbf{M}_{q,k}^{\prime\prime\prime} \mathbf{p} (\mathbf{p}^H \mathbf{M}_{q,k}^{\prime\prime\prime} \mathbf{p} )^H$}.
Then, the simplified weighted sum-$v_\text{out}$ maximization problem $\max_{\mathbf{p}} \{\sum_{q=1}^K w_q v_{\text{out},q}(\mathbf{p}) : \|\mathbf{p}\|^2 \leq P \}$ can be recast as
\begin{IEEEeqnarray}{cl}
\label{Epi_Problem_SimpMaxMinVout}
\min_{\gamma_1^{\prime\prime}, \{\mathbf{t}_q\}_{q=1}^K, \mathbf{X}\succeq 0} \, & \gamma_1^{\prime\prime} \IEEEyesnumber\IEEEyessubnumber\\
\text{s.t.} & \textstyle{\sum_{q\!=\!1}^K} w_q \! \left(- \! \beta_2 t_{q,0}\! +\! g_q(\mathbf{t}_q)\right) \! - \! \gamma_1^{\prime\prime} \! \leq \! 0,\quad \IEEEyessubnumber \\
& \text{Tr}\{\mathbf{M}_{q,k}^{\prime\prime\prime} \mathbf{X}\} = t_{q,k}\,,\, \forall q, k\,, \IEEEyessubnumber \\
& \text{Tr}\{[\mathbf{M}_{q,k}^{\prime\prime\prime}]^H \mathbf{X}\} = t_{q,k}^\ast\,,\, \forall q, k\neq0\,, \IEEEyessubnumber \\
& \text{Tr}\{\mathbf{X}\} \leq P\,, \IEEEyessubnumber\\
& \text{rank}\{\mathbf{X}\} = 1\,.\IEEEyessubnumber
\end{IEEEeqnarray}

It is shown that problem (\ref{Epi_Problem_SimpMaxMinVout}) has the same structure as problem (\ref{Epi_Problem_MaxVout}). Hence, (\ref{Epi_Problem_SimpMaxMinVout}) can be solved by a variation of Algorithm \ref{AlgSCA}. In order to apply Algorithm \ref{AlgSCA} to (\ref{Epi_Problem_SimpMaxMinVout}), changes are made as follows. The matrix $\mathbf{A}_1$ in Line \ref{AlgSCALine4} of Algorithm \ref{AlgSCA} should be replaced by $\mathbf{A}_1^{\prime\prime} \triangleq \mathbf{C}_1^{\prime\prime} + [\mathbf{C}_1^{\prime\prime}]^H$, where
{\small$\mathbf{C}_1^{\prime\prime} \! = \! \sum_{q = 1}^K \! w_q \big(\! -\frac{\beta_2  +  3\beta_4t^{(l\!-\!1)}_{q,0}}{2}\mathbf{M}_{q,0}^{\prime\prime\prime} \! - \! 3\beta_4 \! \sum_{k=1}^{N\!-\!1} [t^{(l \!- \!1)}_{q,k}]^\ast \mathbf{M}_{q,k}^{\prime\prime\prime} \! \big)$}.
Line \ref{AlgSCALine4} of Algorithm \ref{AlgSCA} should be modified as $\mathbf{p}^\star = \mathbf{x}^\star$ and $\mathbf{s}_n^\star = [\mathbf{p}^\star]_{n,1} \mathbf{w}_n \forall n$. Due to space constraint, the algorithm is not outlined in pseudocode as Algorithm \ref{AlgSCA}.

\subsection{Exploiting Channel Hardening}\label{SecAsympAnaWSum}
This subsection proposes a Channel Hardening-Exploiting (CHE) weighted sum $v_\text{out}$ maximization algorithm, under the assumption that the channel of a given user is sufficiently frequency-selective such that channel gains can be i.i.d. in space and frequency. Additionally, the channels across users are fully uncorrelated. That is, given the large-scale fading $\Lambda^{1/2}_q$ and $h_{q,(n-1)M+1} \! \sim \! \mathcal{CN}(0, \Lambda_q)$, channel hardening indicates that as $M\! \rightarrow\! \infty$, $\mathbf{h}_{q,n}^T \mathbf{h}_{q,n}^\ast/M\! = \! \Lambda_q$ and $\mathbf{h}_{q,n}^T \mathbf{h}_{q^\prime,n^\prime}^\ast/M\! =\! 0$ for $q^\prime \!\neq\! q$ or $n^\prime \!\neq \! n$.

\subsubsection{Asymptotical Analysis}
As defined in Section \ref{SecSignalTrans}, the spatial domain precoders across frequencies are collected in $\mathbf{s}$. In the following, the asymptotically optimal $\mathbf{s}$ is referred to as $\mathbf{s}_\text{asym}$, of which the normalized version is designated as $\mathbf{\bar{s}} = [\mathbf{\bar{s}}_1^T, \ldots, \mathbf{\bar{s}}_N^T]^T$. Hence, $\mathbf{\bar{s}}_n$ is subject to $\sum_{n = 1}^N \|\mathbf{\bar{s}}_n\|^2 = 1$. Then, the optimal structure of $\mathbf{\bar{s}}_n$ can be
\begin{equation}
\label{EqSnBar}
\mathbf{\bar{s}}_n = \textstyle{\sum_{q = 1}^K} \xi_{q,n}\mathbf{h}_{q,n}^\ast/\sqrt{M}\,,
\end{equation}
where $\xi_{q,n}$ is a complex weight. With such a $\mathbf{\bar{s}}_n$, by defining $E \triangleq PM$, the asymptotically optimal\footnote{The optimality of $\mathbf{\bar{s}}_n$ can be shown by contradiction as in\cite{XTW14}. Intuitively, if $\mathbf{\bar{s}}_n$ contains not only the linear combination of $\xi_{q,n}\mathbf{h}_{q,n}^\ast$ as in (\ref{EqSnBar}) but also the directions in the orthogonal complement of the space spanned by $\{\mathbf{h}_{q,n}^\ast\}_{q=1}^K$, some transmit power would not contribute to $v_\text{out}$ and is wasted.} $\mathbf{s}$ can be written as
\begin{equation}
\label{EqSasym}
\mathbf{s}_\text{asym} \triangleq \sqrt{E/M}\mathbf{\bar{s}}\,.
\end{equation}
Substituting $\mathbf{s}_\text{asym}$ into (\ref{EqFuncVoutq}) and exploiting channel hardening, the asymptotic output voltage at user $q$ can be written as
\begin{IEEEeqnarray}{rcl}
\label{EqAsymptOutputVol}
v_{\text{out},q}^\prime &{}={}& \beta_2 E\Lambda_q^2\mathbf{p}_q^H\mathbf{p}_q \! + \! \frac{3}{2}\beta_4 E^2\Lambda_q^4 \! \left(\!\mathbf{p}_q^H \mathbf{M}_0^\prime \mathbf{p}_q \! \right)\!\left(\mathbf{p}_q^H \mathbf{M}_0^\prime \mathbf{p}_q \! \right)^H \nonumber\\
&&{}+{}3\beta_4 E^2\Lambda_q^4 \textstyle{\sum_{k=1}^{N-1}}\! \left(\!\mathbf{p}_q^H \mathbf{M}_k^\prime \mathbf{p}_q \! \right)\!\left(\mathbf{p}_q^H \mathbf{M}_k^\prime \mathbf{p}_q \! \right)^H\,,
\end{IEEEeqnarray}
where $\mathbf{p}_q \! = \! [\xi_{q,1}\!, \ldots\!, \xi_{q,N}]^T$. In (\ref{EqAsymptOutputVol}), $\mathbf{M}^\prime_k$ returns a $N$-by-$N$ matrix whose $k$\,th diagonal\footnote{For $k \!> \!0$, $k$ is the index of the $k$\,th diagonal above the main diagonal (whose index $k=0$); for $k\!<\!0$, $k$ is the index of the $|k|$\,th diagonal below the main diagonal. For instance, $\mathbf{M}^\prime_0 = \mathbf{I}_{N\times N}$.} is made of ones, while all the other entries are zero. Assuming uniform power allocation across frequencies and considering the constraint on $\mathbf{\bar{s}}_n$, $\mathbf{p}_1$ can be written as $\mathbf{p}_1 = \frac{1} {\sqrt{N\Lambda_1}} \mathbf{1}_{N\times 1}$. Then, (\ref{EqAsymptOutputVol}) becomes
\begingroup\makeatletter\def\f@size{9}\check@mathfonts
\def\maketag@@@#1{\hbox{\m@th\small\normalfont#1}}%
\begin{equation}
\label{EqAsymptOutputVol_UniformPwrAlloc}
v_{\text{out},1}^\prime \! = \! \beta_2 E \Lambda_1 \! + \! 3\beta_4 E^2\Lambda_1^2/2 \!+\! \beta_4 E^2\Lambda_1^2 N(\!N\!-\!1)(2N\!-1)/(2N^2)\,.
\end{equation}
\endgroup
It is shown that if $N$ is sufficiently large, $v_{\text{out},q}^\prime$ linearly scales with $N$. This observation is in line with the scaling law of \cite{CBmar16TSParxiv} for frequency-selective channels.

\subsubsection{Asymptotically Optimal Waveform Design}
The weighted sum $v_\text{out}$ maximization problem can be formulated as
\begin{IEEEeqnarray}{cl}
\label{OriginalProblem_AsymptWsum}
\max_{\{\mathbf{p}_q\}_{q=1}^K} \quad & \left\{ \textstyle{\sum_{q=1}^K} w_q \cdot v_{\text{out},q}^\prime: \sum_{q=1}^K \Lambda_q \|\mathbf{p}_q\|^2 = 1 \right\}\,.
\end{IEEEeqnarray}
Therefore, solving problem (\ref{OriginalProblem_AsymptWsum}) essentially optimizes $\xi_{q,n}$, which is related to the frequency domain power allocation, relying on large-scale fading. However, (\ref{EqSnBar}) implies that to conduct WPT, the waveform precoder at each frequency would be a function of $\mathbf{h}_{q,n}$, relying on short-term CSI.

Similarly to Section \ref{SecGeneWirelessChan}, in order to solve (\ref{OriginalProblem_AsymptWsum}), it is reformulated as
\begin{IEEEeqnarray}{l}
\label{EquiProblem_AsymptWsum}
\min_{\gamma_1^\prime\!, \{\!\mathbf{p}_q\!\}_{q\!=\!1}^K\!, \{\!\mathbf{t}_q\!\}_{q\!=\!1}^K} \quad \gamma_1^\prime\IEEEyesnumber\IEEEyessubnumber\\
\text{s.t.} \quad \textstyle{\sum_{q=1}^K} \! w_q \! \left(\!E^2 \Lambda_q^4 g_q(\mathbf{t}_q) \! - \!\beta_2 E\Lambda_q^2t_{q,0} \!\right) \! \leq \! \gamma_1^\prime, \IEEEyessubnumber \label{EqNoncovxConst_EquiProbAsymptWsum}\\
\quad\quad\, \mathbf{p}_q^H \mathbf{M}^\prime_k \mathbf{p}_q = t_{q,k}\,, \forall q,k \IEEEyessubnumber \label{EqConst2_EquiProbAsymptWsum}\\
\quad\quad\, \mathbf{p}_q^H \left[\mathbf{M}^\prime_k\right]^H \mathbf{p}_q = t_{q,k}^\ast\,, \forall q,k\neq 0 \IEEEyessubnumber \label{EqConst3_EquiProbAsymptWsum}\\
\quad\quad\, \textstyle{\sum_{q=1}^K} \Lambda_q \|\mathbf{p}_q\|^2 = 1\,, \IEEEyessubnumber \label{EqConstNorml_EquiProbAsymptWsum}
\end{IEEEeqnarray}
where $k \!=\! 0,\ldots,N\!-\!1$. To solve the problem, $g_q(\mathbf{t}_q)$ in (\ref{EqNoncovxConst_EquiProbAsymptWsum}) is linearized by (\ref{Eq_tilde_g_q}), yielding the AP of (\ref{EquiProblem_AsymptWsum}) at the $l$\,th iteration
\begin{IEEEeqnarray}{l}
\label{ApproxProblem_AsymptWsum}
\min_{\gamma_1^\prime\!, \{\!\mathbf{p}_q\!\}_{q=1}^K\!, \{\!\mathbf{t}_q\!\}_{q=1}^K} \,  \gamma_1^\prime\IEEEyesnumber\IEEEyessubnumber\\
\text{s.t.} \, \textstyle{\sum_{q=1}^K}\! w_q \! \left(\!- \!\beta_2 E\!\Lambda_q^2 t_{q,0} \!+\! E^2\!\Lambda_q^4\tilde{g}_q\!(\mathbf{t}_q, \mathbf{t}_q^{(l \! - \! 1)})  \!\right) \!\leq \! \gamma_1^\prime, \quad \IEEEyessubnumber\\
\quad \quad\text{(\ref{EqConst2_EquiProbAsymptWsum}), (\ref{EqConst3_EquiProbAsymptWsum}) and (\ref{EqConstNorml_EquiProbAsymptWsum})} \nonumber\,.
\end{IEEEeqnarray}
Different from the APs (\ref{ApproxConvProblem_MaxVoutSU}) and (\ref{ApproxConvProblem_MaxVout}), the AP (\ref{ApproxProblem_AsymptWsum}) is nonconvex.
Fortunately, the global optimum of (\ref{ApproxProblem_AsymptWsum}) can be achieved by solving an equivalent problem. Substitute (\ref{EqConst2_EquiProbAsymptWsum}) and (\ref{EqConst3_EquiProbAsymptWsum}) into (\ref{EqNoncovxConst_EquiProbAsymptWsum}). By defining
\begin{equation}
\label{EqCq1prime}
\mathbf{C}_{\!q,1}^\prime\! \triangleq\! -\!\frac{\beta_2\! E\! \Lambda_q^2\! +\! 3\! E^2\! \Lambda_q^4\beta_{4} t_{\!q,0}^{(l\! -\! 1)}}{2}\mathbf{M}^\prime_0\! -\! 3\beta_{4}\! E^2\! \Lambda_q^4\! \sum_{k=1}^{N\! -\! 1} \! \left[t_{\!q,k}^{(l\! -\! 1)}\right]^\ast\! \mathbf{M}^\prime_k
\end{equation}
and
\begin{equation}
\label{EqAq1prime}
\mathbf{A}^\prime_{q,1} \triangleq \mathbf{C}_{q,1}^\prime + [\mathbf{C}_{q,1}^\prime]^H\,,
\end{equation}
the equivalent form of (\ref{ApproxProblem_AsymptWsum}) can be formulated as
\begin{IEEEeqnarray}{cl}
\label{EquiProb_Approx_AsymptWsum}
\min_{\mathbf{\bar{p}}} \quad & \left\{ \mathbf{\bar{p}}^H \mathbf{A}^\prime_1 \mathbf{\bar{p}}: \mathbf{\bar{p}}^H \mathbf{\Lambda}\mathbf{\bar{p}} = 1 \right\}\,,
\end{IEEEeqnarray}
where {\small$\mathbf{\bar{p}}\! \triangleq\! [\mathbf{p}_1^T, \ldots, \mathbf{p}_K^T]^T$}, {\small$\mathbf{A}^\prime_1 \!\triangleq \! diag\{w_1\mathbf{A}^\prime_{1,1}, \ldots, w_K\mathbf{A}^\prime_{K,1}\}$} and {\small$\mathbf{\Lambda}\!\triangleq \!diag\{\mathbf{\Lambda}_1, \ldots, \mathbf{\Lambda}_K\}$}. All the main diagonal entries of the $N$-by-$N$ diagonal matrix {\small$\mathbf{\Lambda}_q$} are equal to {\small$\Lambda_q$}. Let {\small$\mathbf{U}_{\mathbf{\bar{\Lambda}}}$} collect the eigenvectors of {\small$\mathbf{\bar{\Lambda}} \!\triangleq\! \mathbf{\Lambda}^{-1}\! \mathbf{A}^\prime_1$}. Similarly to Proposition \ref{PropRank1soluSU}, the optimal {\small$\mathbf{\bar{p}}^\star \! = \! b \big[ \mathbf{U}_{\mathbf{\bar{\Lambda}}} \big]_\text{min}$}, where {\small$b \! = \! \big[ 1/ \big(\big[\mathbf{U}_{\mathbf{\bar{\Lambda}}} \big]_\text{min}^H \mathbf{\Lambda} \big[\mathbf{U}_{\mathbf{\bar{\Lambda}}} \big]_\text{min} \big) \big]^{1/2}$}. For {\small$\Lambda_q \!=\! \Lambda \,\, \forall q$}, {\small$\mathbf{\bar{p}}^\star \!=\! \sqrt{\frac{1}{\Lambda}} \big[ \mathbf{U}_{\mathbf{A}^\prime_1} \big]_\text{min}$}, where {\small$\mathbf{U}_{\mathbf{A}^\prime_1}$} collects the eigenvectors of {\small$\mathbf{A}^\prime_1$}. Note that the EVD of {\small$\mathbf{\bar{\Lambda}}$} (or {\small$\mathbf{A}^\prime_1$}) yields complexity of $O\big(KN^3\big)$.
\begin{algorithm}
\caption{CHE WSum Algorithm}\label{AlgAsymptWsum}
\begin{algorithmic}[1]
\State \textbf{Initialize} Set $l = 0$, and generate feasible initial points $\{\mathbf{p}_q^{(0)}\}_{q=1}^K$. Then, compute $\{\mathbf{t}_q^{(0)}\}_{q = 1}^K$ by (\ref{EqConst2_EquiProbAsymptWsum}).
\Repeat
    \State $l = l + 1$;
    \State Compute $\{\mathbf{C}_{q,1}^\prime(\mathbf{t}_q^{(l-1)})\}_{q=1}^K$ and $\{\mathbf{A}^\prime_{q,1}(\mathbf{t}_q^{(l-1)})\}_{q=1}^K$;
    \State Compute $\mathbf{A}^\prime_1$, $\mathbf{\bar{p}}^\star = b \big[ \mathbf{U}_{\mathbf{\bar{\Lambda}}} \big]_\text{min}$ and $\{\mathbf{p}_q^\star\}_{q=1}^K$;
    \State Update $\mathbf{\bar{p}}^{(l)} = \mathbf{\bar{p}}^\star$ and $\mathbf{p}_q^{(l)} = \mathbf{p}_q^\star, \forall q$; Update $t_{q,k}^{(l)}$;
\Until{\|\mathbf{\bar{p}}^{(l)} - \mathbf{\bar{p}}^{(l-1)}\|/\|\mathbf{\bar{p}}^{(l)}\| \leq \epsilon}
\State $\mathbf{\bar{s}}_n^\star = \sum_{q = 1}^K [\mathbf{p}_q^\star]_{n,1} \mathbf{h}_{q,n}^\ast/\sqrt{M}$ and $\mathbf{s}_\text{asym}^\star = \sqrt{E/M}\mathbf{\bar{s}}^\star/\|\mathbf{\bar{s}}^\star\|$ \label{AlgoLine_CHE_WSum_wfPrecoder}
\end{algorithmic}
\end{algorithm}
The algorithm is summarized in Algorithm \ref{AlgAsymptWsum}. As (\ref{EquiProb_Approx_AsymptWsum}) yields the global optimum of (\ref{ApproxProblem_AsymptWsum}) and $\tilde{g}_q(\mathbf{t}_q^{(l)}\!, \mathbf{t}_q^{(l)})\! =\! g_q(\mathbf{t}_q^{(l)}) \!\leq\! \tilde{g}_q(\mathbf{t}_q^{(l)}\!, \mathbf{t}_q^{(l-1)})$, the objective function of (\ref{ApproxProblem_AsymptWsum}) decreases over iterations. Thus, Algorithm \ref{AlgAsymptWsum} finally converges to a stationary point of the original problem (\ref{EquiProblem_AsymptWsum}) provided the eigenvectors of a given $\mathbf{\bar{\Lambda}}$ are uniquely attained.

\emph{Remark:} Line \ref{AlgoLine_CHE_WSum_wfPrecoder} in Algorithm \ref{AlgAsymptWsum} indicates that for $K=1$, the waveform precoder at frequency $n$ (generated by Algorithm \ref{AlgAsymptWsum}) is $\mathbf{s}_n = \frac{\sqrt{P} \xi_{1,n}^\star \mathbf{h}_{1,n}^\ast}{\sqrt{\sum |\xi_{1,n}^\star|^2 \|\mathbf{h}_{1,n}\|^2}}$, implying that each frequency component may still be allocated a different power. This is different from the waveform strategy used for deriving the scaling law in \cite{CBmar16TSParxiv}. Recall that to derive the scaling law, \cite{CBmar16TSParxiv} assumes a simple strategy, which performs uniform power (UP) in the frequency domain and matched beamforming (MF), or known as MRT, in the spatial domain, i.e. $\mathbf{s}_n = \sqrt{P/N} \frac{\mathbf{h}_{1,n}^\ast}{\|\mathbf{h}_{1,n}\|}$.

\section{Minimum $v_\text{out}$ Maximization Algorithms}\label{SecMaxMinAlgos}
\subsection{Minimum $v_\text{out}$ Maximization}
This subsection derives fairness-aware $K$-user waveform optimization algorithms, by maximizing the minimum $v_\text{out}$:
\begin{equation}
\label{Problem_MaxMinEquivalent}
\max_{\mathbf{s}, \gamma_2} \{\gamma_2: v_{\text{out},q}(\mathbf{s}) \geq \gamma_2\, \forall q, \|\mathbf{s}\|^2 \leq P\}\,.
\end{equation}
For $K \!=\! 1$, (\ref{Problem_MaxMinEquivalent}) boils down to (\ref{ProbOriginalSUwpt}). To address the quartic polynomial $v_{\text{out},q}(\mathbf{s})$, auxiliary variables $t_{q,k}$ are introduced as in Section \ref{SecGeneWirelessChan}, such that (\ref{Problem_MaxMinEquivalent}) can be equivalently recast as
\begin{IEEEeqnarray}{cl}
\label{Prob_MaxMinEqui_lowDeg}
\min_{\gamma_2,\, \{\mathbf{t}_q\}_{q=1}^K,\, \mathbf{X}\succeq 0} \quad & - \gamma_2 \IEEEyesnumber\IEEEyessubnumber\\
\text{s.t.} &  -\! \beta_2 t_{q,0}\! +\! g_q(\mathbf{t}_q)\! +\! \gamma_2\! \leq\! 0 \,,\,\forall q\,, \IEEEyessubnumber \label{Eq_NonCVX_Prob_MaxMinEqui_lowDeg}\\
& \text{rank}\{\mathbf{X}\} = 1\,.\IEEEyessubnumber\\
& \text{(\ref{EqConst_t_qk}), (\ref{EqConstConjugatet_qk}) and (\ref{EqTxPwrConst})}\,.\nonumber
\end{IEEEeqnarray}
Similarly to (\ref{Epi_Problem_MaxVout}), (\ref{Prob_MaxMinEqui_lowDeg}) is nonconvex, due to the nonconvex (\ref{Eq_NonCVX_Prob_MaxMinEqui_lowDeg}) and the rank constraint. To solve (\ref{Prob_MaxMinEqui_lowDeg}), we first relax the rank constraint, yielding
\begin{IEEEeqnarray}{cl}
\label{Prob_MaxMinEqui_RankRelax}
\min_{\gamma_2, \{\mathbf{t}_q\}_{q=1}^K, \mathbf{X}\succeq 0} \quad & - \gamma_2 \IEEEyesnumber\IEEEyessubnumber\\
\text{s.t.} \,&  - \beta_2 t_{q,0} + g_q(\mathbf{t}_q) + \gamma_2 \leq 0 \,,\forall q, \IEEEyessubnumber\label{Eq_NonCVX_Prob_MaxMinEqui_RankRelax}\\
& \text{(\ref{EqConst_t_qk}), (\ref{EqConstConjugatet_qk}) and (\ref{EqTxPwrConst})}\,.\nonumber
\end{IEEEeqnarray}
Therefore, achieving the solution of (\ref{Prob_MaxMinEqui_lowDeg}) amounts to finding a rank-constrained solution by solving (\ref{Prob_MaxMinEqui_RankRelax}). Problem (\ref{Prob_MaxMinEqui_RankRelax}) is nonconvex due to (\ref{Eq_NonCVX_Prob_MaxMinEqui_RankRelax}). Therefore, we approximate the solution of (\ref{Prob_MaxMinEqui_RankRelax}) by SCA over iterations as in Section \ref{SecGeneWirelessChan}. Hence, by approximating (or linearizing) $g_q(\mathbf{t}_q)$ by (\ref{Eq_tilde_g_q}), the AP of (\ref{Prob_MaxMinEqui_RankRelax}) at iteration $l$ can be formulated as
\begin{IEEEeqnarray}{cl}
\label{ApproxConvProblem_MaxMinEqui}
\min_{\gamma_2, \{\mathbf{t}_q\}_{q\!=\!1}^K, \mathbf{X}\succeq 0}  & - \gamma_2 \IEEEyesnumber\IEEEyessubnumber \label{EqApprox_ApproxConvProblem_MaxMinEqui}\\
\text{s.t.} & -\! \beta_2 t_{q,0} \! +\! \tilde{g}_q\!\left(\mathbf{t}_q, \mathbf{t}_q^{(l \! - \! 1)}\right)\! +\!\gamma_2 \leq 0 \,, \forall q,\quad \IEEEyessubnumber\\
& \text{(\ref{EqConst_t_qk}), (\ref{EqConstConjugatet_qk}) and (\ref{EqTxPwrConst})}.\nonumber
\end{IEEEeqnarray}

\subsubsection{Max-Min Algorithm with Rank Reduction (for $K \leq 3$)}\label{SecMaxMin_RR}
We first show that if $K \leq 3$, (\ref{ApproxConvProblem_MaxMinEqui}) can offer a solution with a rank-1 optimal $\mathbf{X}_\text{r1}$, although solving (\ref{ApproxConvProblem_MaxMinEqui}) by the interior-point method usually yields a solution with a \emph{high-rank} $\mathbf{X}^\star$\cite{Dattorro05}. Based on this, in order to solve (\ref{Prob_MaxMinEqui_lowDeg}), we then propose an algorithm based on SCA and a embedded Rank Reduction (RR) procedure, where the RR procedure is applied to the high-rank solution of the AP (\ref{ApproxConvProblem_MaxMinEqui}) to achieve a rank-1 $\mathbf{X}_\text{r1}$.
\begin{theorem}
\label{TheoKleq3_MaxMinRank1}
Given $K \leq 3$, problem (\ref{ApproxConvProblem_MaxMinEqui}) can yield, among others, an optimal $\mathbf{X}^\star$ of rank 1.
\end{theorem}
\begin{IEEEproof}
Substituting (\ref{EqConst_t_qk}) and (\ref{EqConstConjugatet_qk}) into (\ref{EqApprox_ApproxConvProblem_MaxMinEqui}) and defining {\small$\bar{c}_q = - [\mathbf{t}_q^{(l - 1)}]^H \mathbf{A}_0 \mathbf{t}_q^{(l - 1)}$}, {\small$\mathbf{C}_{q,1} \! = \! -\frac{\beta_2 \! + \!3\beta_4t^{(l \! - \! 1)}_{q,0}}{2}\mathbf{M}_{q,0} - 3\beta_4 \! \sum_{k=1}^{N-1} \! \left[t^{(l-1)}_{q,k}\right]^\ast \! \mathbf{M}_{q,k}$} and {\small$\mathbf{A}_{q,1} = \mathbf{C}_{q,1} + \mathbf{C}_{q,1}^H$}, problem (\ref{ApproxConvProblem_MaxMinEqui}) can be equivalently reformulated as
\begin{equation}
\label{ProbEqui_ApproxConvProblem_MaxMinEqui_1}
\min_{\gamma_2, \mathbf{X}\succeq 0} \{ - \gamma_2: \text{Tr}\{\mathbf{A}_{q,1}\mathbf{X}\}\! +\! \bar{c}_q \!+\! \gamma_2 \!\leq\! 0 \, \forall q, \text{Tr}\{\mathbf{X}\} \!\leq\! P\}\,,
\end{equation}
which can be transformed into another equivalent form:
\begin{equation}
\label{ProbEqui_ApproxConvProblem_MaxMinEqui}
\min_{\mathbf{X}\succeq 0} \max_{q = 1\ldots K} \{\text{Tr}\{\mathbf{A}_{q,1}\mathbf{X}\} + \bar{c}_q: \text{Tr}\{\mathbf{X}\} \leq P\}\,.
\end{equation}
Given the optimal $q^\star =  q_0$ that maximizes $\text{Tr}\{\mathbf{A}_{q,1}\mathbf{X}\} + \bar{c}_q$, problem (\ref{ProbEqui_ApproxConvProblem_MaxMinEqui}) boils down to
\begin{IEEEeqnarray}{cl}
\label{Prob_MaxMinEqui_rankReduct}
\min_{\mathbf{X}\succeq 0}\,  & \text{Tr}\{\mathbf{A}_{q_0,1}\mathbf{X}\} \IEEEyesnumber\IEEEyessubnumber \\
\text{s.t.}\, & \text{Tr}\left\{\left(\mathbf{A}_{q,1}\!-\!\mathbf{A}_{q_0,1}\right)\mathbf{X}\right\}\! \leq\! \bar{c}_{q_0}\! -\! \bar{c}_q, \forall q\! \neq\! q_0, \IEEEyessubnumber\\
& \text{Tr}\{\mathbf{X}\} \leq P\,.\IEEEyessubnumber
\end{IEEEeqnarray}
In the case of $K \leq 3$, (\ref{Prob_MaxMinEqui_rankReduct}) turns out to be a separable SDP, and the number of its linear constraints is no greater than three. It can be shown that (\ref{Prob_MaxMinEqui_rankReduct}) can yield, among others, a rank-1 optimal solution\cite[Proposition 3.5]{HP10}. Therefore, for $K \leq 3$, problem (\ref{ApproxConvProblem_MaxMinEqui}) can yield a rank-1 $\mathbf{X}_\text{r1}$.
\end{IEEEproof}

\begin{algorithm}
\caption{Max-Min-RR Algorithm (for $K \leq 3$)}\label{AlgMaxMinKleq3}
\begin{algorithmic}[1]
\State \textbf{Initialize} Set $l = 0$, and generate $\mathbf{X}^{(0)}$ and $\{\mathbf{t}_q^{(0)}\}_{q = 1}^K$;
\Repeat
    \State $l = l + 1$;
    \State Update $\bar{c}_q$, $\mathbf{C}_{q,1}$ and $\mathbf{A}_{q,1}$; solve problem (\ref{ProbEqui_ApproxConvProblem_MaxMinEqui_1}) by the interior-point method, yielding a high-rank solution $\mathbf{X}^\star$;\label{Line5_AlgMaxMinKleq3}
    \State Find the optimal $q^\star \! \triangleq \! q_0 \! =\! \arg\max_{q} \text{Tr}\{\mathbf{A}_{q,1}\mathbf{X}^\star\} + \bar{c}_q$;\label{Line6_AlgMaxMinKleq3}
    \State Apply the RR procedure \cite[Algorithm 1]{HP10} to problem (\ref{Prob_MaxMinEqui_rankReduct}), yielding a rank-1 optimal solution  $\mathbf{X}_\text{r1} = \mathbf{x}_0\mathbf{x}_0^H$;\label{Line7_AlgMaxMinKleq3}
    \State Update $\mathbf{X}^{(l)} = \mathbf{X}_\text{r1}$; update $t_{q,k}^{(l)}$ $\forall q,k$ by (\ref{EqConst_t_qk});
\Until{\|\mathbf{X}^{(l)} - \mathbf{X}^{(l-1)}\|_F/\|\mathbf{X}^{(l)}\|_F \leq \epsilon}
\State $\mathbf{s}^\star = \mathbf{x}_0$.
\end{algorithmic}
\end{algorithm}
According to Theorem \ref{TheoKleq3_MaxMinRank1}, for $K \leq 3$, Algorithm \ref{AlgMaxMinKleq3} is proposed to solve (\ref{Prob_MaxMinEqui_lowDeg}). In the algorithm, a rank-1 solution $\mathbf{X}_\text{r1}$ of (\ref{ApproxConvProblem_MaxMinEqui}) can be obtained by performing RR, such that solving (\ref{ApproxConvProblem_MaxMinEqui}) iteratively can finally make the solution $(\gamma_2, \{\mathbf{t}_q\}_{q\!=\!1}^K, \mathbf{X}_\text{r1})$ converge to a stationary point of (\ref{Prob_MaxMinEqui_lowDeg}). The proposed algorithm is elaborated and explained as follows.

Exploiting the interior-point method\footnote{In our simulations, SDPs are solved by CVX. CVX invokes the universal solver SDPT3 to implement the interior-point algorithm\cite{GB14}.} to solve (\ref{ProbEqui_ApproxConvProblem_MaxMinEqui_1}), Line \ref{Line5_AlgMaxMinKleq3} of Algorithm \ref{AlgMaxMinKleq3} essentially solves the $l$\,th AP (\ref{ApproxConvProblem_MaxMinEqui}). As $\tilde{g}_q(\mathbf{t}_q^{(l)}, \mathbf{t}_q^{(l)}) = g_q(\mathbf{t}_q^{(l)}) \leq \tilde{g}_q(\mathbf{t}_q^{(l)}, \mathbf{t}_q^{(l-1)})$ and $\nabla g_q(\mathbf{t}_q^{(l)}) = \nabla \tilde{g}_q(\mathbf{t}_q^{(l)}, \mathbf{t}_q^{(l)})$, it can be shown that solving (\ref{ApproxConvProblem_MaxMinEqui}) iteratively can achieve a stationary point of (\ref{Prob_MaxMinEqui_RankRelax}). Intuitively, if the optimal $\mathbf{X}^\star$ of (\ref{ApproxConvProblem_MaxMinEqui}) always remains rank-1, the stationary point of (\ref{Prob_MaxMinEqui_RankRelax}) is also that of (\ref{Prob_MaxMinEqui_lowDeg}).
According to Theorem \ref{TheoKleq3_MaxMinRank1}, a RR procedure can be exploited to derive a rank-1 solution $\mathbf{X}_\text{r1}$ from the high-rank solution of (\ref{ApproxConvProblem_MaxMinEqui}). To the end, Line \ref{Line6_AlgMaxMinKleq3} finds the optimal $q^\star$, such that the parameters $\mathbf{A}_{q_0,1}$, $\mathbf{A}_{q,1}\!-\!\mathbf{A}_{q_0,1}$ and $\bar{c}_{q_0}\! -\! \bar{c}_q$ of problem (\ref{Prob_MaxMinEqui_rankReduct}) can be computed. Then, in Line \ref{Line7_AlgMaxMinKleq3}, the RR procedure \cite[Algorithm 1]{HP10} is applied to problem (\ref{Prob_MaxMinEqui_rankReduct}) to obtain a rank-1 $\mathbf{X}_\text{r1}$, according to the proof of Theorem \ref{TheoKleq3_MaxMinRank1}.
Although the RR procedure \cite[Algorithm 1]{HP10} is iterative, the optimal $q^\star$ remains constant, as the values of $\text{Tr}\{\mathbf{X}\}$ and $\text{Tr}\{(\mathbf{A}_{q,1}\!-\!\mathbf{A}_{q_0,1})\mathbf{X}\}$ ($\forall q\! \neq\! q_0$) remain constant over iterations\footnote{This is because the RR formula at Line 8 of \cite[Algorithm 1]{HP10} keeps the values of the constraint functions constant over iterations. See the proof of \cite[Lemma 3.1]{HP10} for details.}. As the RR procedure preserves the primal feasibility and the complementary slackness of (\ref{Prob_MaxMinEqui_rankReduct}), the achieved $\mathbf{X}_\text{r1}$ remains globally optimal for (\ref{Prob_MaxMinEqui_rankReduct}).
Such a RR procedure is deterministic\cite{HP14}, namely, given $\mathbf{A}_{q,1}$, $\bar{c}_q$ and $P$, the optimal rank-1 $\mathbf{X}_\text{r1}$ is uniquely attained. Hence, similarly to Theorem \ref{TheoConvergenceAlgSCA_SU}, we can show that Algorithm \ref{AlgMaxMinKleq3} converges to a stationary point of (\ref{Prob_MaxMinEqui_RankRelax}). As the solution is rank-1, it is also a stationary point of (\ref{Prob_MaxMinEqui_lowDeg}).

\subsubsection{Max-Min Algorithm with Randomization (for an Arbitrary $K$)}\label{SecSCAthenRand}
It can be seen from the proof of Theorem \ref{TheoKleq3_MaxMinRank1} that when $K > 3$, problem (\ref{Prob_MaxMinEqui_rankReduct}) may not have a rank-1 optimal solution. Therefore, (\ref{Prob_MaxMinEqui_RankRelax}) cannot always yield a rank-1 optimal $\mathbf{X}^\star$.
Then, a SCA-then-randomization method is used to solve (\ref{Prob_MaxMinEqui_lowDeg}), such that a rank-1 $\mathbf{X}$ can be achieved for (\ref{Prob_MaxMinEqui_lowDeg}).

\begin{algorithm}
\caption{Max-Min-Rand Algorithm (for an Arbitrary $K$)}\label{AlgMaxMinRand_ArbiK}
\begin{algorithmic}[1]
\State \textbf{Initialize} Set $l = 0$, and generate $\mathbf{X}^{(0)}$ and $\{\mathbf{t}_q^{(0)}\}_{q = 1}^K$;
\Repeat
    \State $l = l + 1$;
    \State Solve (\ref{ApproxConvProblem_MaxMinEqui}) by the interior-point method, yielding $\mathbf{X}^\star$;
    \State Update $\mathbf{X}^{(l)} = \mathbf{X}^\star$; update $t_{q,k}^{(l)}$ $\forall q,k$ by (\ref{EqConst_t_qk});
\Until{\|\mathbf{X}^{(l)} - \mathbf{X}^{(l-1)}\|_F/\|\mathbf{X}^{(l)}\|_F \leq \epsilon}
\State Perform EVD $\mathbf{X}^\star\! =\! \mathbf{U}_X \! \mathbf{\Sigma}_X \! \mathbf{U}_X^H$; for $t = 1,\ldots, T$, generate $\mathbf{\hat{x}}_t \! =\! \mathbf{U}_X \! \mathbf{\Sigma}_X^{1/2} \! \mathbf{v}_t$ with $\mathbf{v}_t$ drawn from a circular uniform distribution;
\State $\mathbf{X}_t\! =\! \mathbf{\hat{x}}_t\mathbf{\hat{x}}_t^H$; the best $\mathbf{X}$ for (\ref{Prob_MaxMinEqui_lowDeg}): $\mathbf{X}_\text{r1}^\star = \arg\min_{\mathbf{X}_t} - \gamma_2(\mathbf{X}_t)$
\end{algorithmic}
\end{algorithm}

Specifically, we exploit SCA to solve (\ref{Prob_MaxMinEqui_RankRelax}). To this end, we formulate the $l$\,th AP of (\ref{Prob_MaxMinEqui_RankRelax}) as (\ref{ApproxConvProblem_MaxMinEqui}). Problem (\ref{ApproxConvProblem_MaxMinEqui}) is iteratively solved by the interior-point method until convergence, yielding a high-rank solution $\mathbf{X}^\star$, which is a stationary point of (\ref{Prob_MaxMinEqui_RankRelax}).
In the following, a best feasible rank-1 solution $\mathbf{X}_\text{r1}^\star$ of (\ref{Prob_MaxMinEqui_lowDeg}) is derived from the high-rank $\mathbf{X}^\star$. To the end, we first perform EVD $\mathbf{X}^\star = \mathbf{U}_X \mathbf{\Sigma}_X \mathbf{U}_X^H$.
Following this, random vectors $\mathbf{\hat{x}}_t$ (for $t = 1, \ldots, T$) are generated: $\mathbf{\hat{x}}_t = \mathbf{U}_X \mathbf{\Sigma}_X^{1/2} \mathbf{v}_t$, where each complex entry in the random vector $\mathbf{v}_t$ is drawn from a circular uniform distribution such that $\mathcal{E}\{\mathbf{v}_t \mathbf{v}_t^H\} = \mathbf{I}$.
Then, $T$ random rank-1 feasible solutions of (\ref{Prob_MaxMinEqui_RankRelax}) can be obtained by $\mathbf{X}_t = \mathbf{\hat{x}}_t\mathbf{\hat{x}}_t^H$. Thanks to the method for generating $\mathbf{\hat{x}}_t$, $\mathbf{X}_t$ always satisfies the power constraint (\ref{EqTxPwrConst}).
As $- \gamma_2$ is the objective function of (\ref{Prob_MaxMinEqui_lowDeg}) and thereby essentially a function of $\mathbf{X}$, we designate $- \gamma_2$ as $- \gamma_2(\mathbf{X})$. Therefore, the best feasible solution (in terms of $\mathbf{X}$) of (\ref{Prob_MaxMinEqui_lowDeg}) is $\mathbf{X}_\text{r1}^\star = \arg\min_{\mathbf{X}_t} - \gamma_2(\mathbf{X}_t)$. The algorithm is summarized in Algorithm \ref{AlgMaxMinRand_ArbiK}.

\subsection{Exploiting Channel Hardening}\label{SecCHE_MaxMinAlgos}
Under the same assumption as in Section \ref{SecAsympAnaWSum}, this subsection proposes simplified algorithm for the minimum $v_\text{out}$ maximization problem, by exploiting channel hardening. It can be shown that the structures of $\mathbf{\bar{s}}_n$ and $\mathbf{s}_\text{asym}$, i.e. (\ref{EqSnBar}) and (\ref{EqSasym}), are still asymptotically optimal. With the asymptotic output voltage (\ref{EqAsymptOutputVol}), the minimum $v_\text{out}$ maximization problem can be formulated as $\max_{\{\mathbf{p}_q\}_{q=1}^K} \min_{q = 1\ldots K} \{ v_{\text{out},q}^\prime: \textstyle{\sum_{q=1}^K} \Lambda_q \|\mathbf{p}_q\|^2 = 1 \}$, which can be equivalently recast as
\begin{IEEEeqnarray}{l}
\label{EquiProblem_AsymptMaxMin}
\min_{\gamma_2^\prime, \{\!\mathbf{p}_q\!\}_{q\!=\!1}^K, \{\!\mathbf{t}_q\!\}_{q\!=\!1}^K} \quad  -\gamma_2^\prime\IEEEyesnumber\IEEEyessubnumber\\
\text{s.t.} \quad E^2 \Lambda_q^4 g_q(\mathbf{t}_q) \! - \!\beta_2 E\Lambda_q^2t_{q,0} \! + \! \gamma_2^\prime \leq \! 0, \forall q \IEEEyessubnumber \label{EqNoncovxConst_EquiProbAsymptMaxMin}\\
\quad \quad\text{(\ref{EqConst2_EquiProbAsymptWsum}), (\ref{EqConst3_EquiProbAsymptWsum}) and (\ref{EqConstNorml_EquiProbAsymptWsum})}\,. \nonumber
\end{IEEEeqnarray}
To solve (\ref{EquiProblem_AsymptMaxMin}), similarly to (\ref{EquiProblem_AsymptWsum}), $g_q(\mathbf{t}_q)$ in (\ref{EqNoncovxConst_EquiProbAsymptMaxMin}) is linearized by (\ref{Eq_tilde_g_q}), such that the $l$\,th AP of (\ref{EquiProblem_AsymptMaxMin}) can be formulated. Reusing the definitions (\ref{EqCq1prime}) and (\ref{EqAq1prime}), and defining
\begin{equation}
\label{EqCqPrime}
\bar{c}_q^\prime = -[\mathbf{t}_q^{(l - 1)}]^H \mathbf{A}_0 \mathbf{t}_q^{(l - 1)} E^2 \Lambda_q^4\,,
\end{equation}
the $l$\,th AP of (\ref{EquiProblem_AsymptMaxMin}) can be written as
\begin{IEEEeqnarray}{cl}
\label{EquiProb_Approx_AsymptMaxMin}
\min_{\gamma_2^\prime, \{\mathbf{p}_q\}_{q=1}^K} \quad & - \gamma_2^\prime \IEEEyesnumber\IEEEyessubnumber\\
\text{s.t.} &  \mathbf{p}_q^H \mathbf{A}_{q,1}^\prime \mathbf{p}_q + \bar{c}_q^\prime + \gamma_2^\prime \leq 0\,,\forall q \IEEEyessubnumber \\
& \textstyle{\sum_{q=1}^K} \Lambda_q \|\mathbf{p}_q\|^2 = 1\,,\IEEEyessubnumber
\end{IEEEeqnarray}
which is nonconvex. By introducing optimization variables $\mathbf{X}_q \triangleq \mathbf{p}_q\mathbf{p}_q^H \forall q$, a Semidefinite Relaxation (SDR) problem, where the constraints $\text{rank}\{\mathbf{X}_q\}\! =\! 1$ for $q = 1,\ldots, K$ have been relaxed, can be formulated as
\begin{IEEEeqnarray}{cl}
\label{EquiProb_Approx_AsymptMaxMinSDR}
\min_{\gamma_2^\prime, \{\mathbf{X}_q \succeq 0\}_{q=1}^K} \quad & - \gamma_2^\prime \IEEEyesnumber\IEEEyessubnumber\\
\text{s.t.} &  \text{Tr}\{\mathbf{A}_{q,1}^\prime \mathbf{X}_q\} + \bar{c}_q^\prime + \gamma_2^\prime \leq 0\,,\forall q, \IEEEyessubnumber \\
& \textstyle{\sum_{q=1}^K} \text{Tr}\{\Lambda_q\mathbf{I}\cdot \mathbf{X}_q\} = 1 \IEEEyessubnumber
\end{IEEEeqnarray}
or
\begin{equation}
\label{ProbEqui_Approx_AsymptMaxMinSDREqui}
\min_{\{\mathbf{X}_q \succeq 0\}_{q=1}^K} \max_{q = 1\ldots K} \{\text{Tr} \{\! \mathbf{A}_{q,1}^\prime \mathbf{X}_q\! \} \!+\! \bar{c}_q^\prime \!: \textstyle{\sum_{q\! =\! 1}^K}\! \text{Tr}\{\! \Lambda_q\mathbf{I}\cdot \mathbf{X}_q \!\}\! =\! 1\}.
\end{equation}
Hence, solving (\ref{EquiProb_Approx_AsymptMaxMin}) boils down to obtaining the rank-constrained optimal $\{\mathbf{X}_q^\star\}_{q=1}^K$ from (\ref{EquiProb_Approx_AsymptMaxMinSDR}) or (\ref{ProbEqui_Approx_AsymptMaxMinSDREqui}).
\begin{theorem}
\label{Theo_CHE_MaxMinRank1}
Problem (\ref{EquiProb_Approx_AsymptMaxMinSDR}) can yield rank-1 optimal $\mathbf{X}_q$ in the presence of an arbitrary number of users.
\end{theorem}
\begin{IEEEproof}
The proof strategy is similar to that for Theorem \ref{TheoKleq3_MaxMinRank1}. Given the optimal $q^\star =  q_0$ that maximizes $\text{Tr} \{\! \mathbf{A}_{q,1}^\prime \mathbf{X}_q\! \} \!+\! \bar{c}_q^\prime$, (\ref{ProbEqui_Approx_AsymptMaxMinSDREqui}) can be reduced to
\begin{IEEEeqnarray}{cl}
\label{Prob_AsymptMaxMinEqui_rankReduct}
\min_{\{\mathbf{X}_q \succeq 0\}_{q=1}^K}\,  & \text{Tr}\{\mathbf{A}_{q_0,1}^\prime\mathbf{X}_{q_0}\} + \bar{c}_{q_0}^\prime \IEEEyesnumber\IEEEyessubnumber \\
\text{s.t.}\, & \text{Tr}\!\left\{\!\left(\mathbf{A}_{\!q,1}^\prime\!-\!\mathbf{A}_{\!q_0,1}^\prime\right)\! \mathbf{X}_{\!q}\! \right\}\! \leq\! \bar{c}_{\!q_0}^\prime\! -\! \bar{c}_{\!q}^\prime, \forall q\! \neq\! q_0, \IEEEyessubnumber\\
& \textstyle{\sum_{q=1}^K} \text{Tr}\{\Lambda_q\mathbf{I}\cdot \mathbf{X}_q\} = 1\,.\IEEEyessubnumber
\end{IEEEeqnarray}
Problem (\ref{Prob_AsymptMaxMinEqui_rankReduct}) contains $K$ linear constraints and $K$ matrix variables. According to \cite[Theorem 3.2]{HP10}, the optimal solutions $\mathbf{X}_q$ of (\ref{Prob_AsymptMaxMinEqui_rankReduct}) satisfy $\sum_{q=1}^K \text{rank}^2\{\mathbf{X}_q^\star\} \leq K$. Therefore, problem (\ref{EquiProb_Approx_AsymptMaxMinSDR}) can yield rank-1 optimal $\mathbf{X}_q$'s for an arbitrary $K$.
\end{IEEEproof}

\subsubsection{CHE Max-Min Algorithm with Rank Reduction (RR)}\label{SecCHeMaxMinRR}
In order to solve (\ref{EquiProb_Approx_AsymptMaxMin}) iteratively and finally achieve a stationary point of (\ref{EquiProblem_AsymptMaxMin}), we propose an algorithm similar to Algorithm \ref{AlgMaxMinKleq3}, where the RR procedure is exploited to obtain the rank-1 optimal solution $\mathbf{X}_q$ for (\ref{EquiProb_Approx_AsymptMaxMinSDR}). Specifically, at each iteration $l$, $\mathbf{A}_{q,1}^\prime$ and $\bar{c}_q^\prime$ are updated for the $l$\,th AP (\ref{EquiProb_Approx_AsymptMaxMinSDR}). Exploiting the interior-point algorithm to solve (\ref{EquiProb_Approx_AsymptMaxMinSDR}) achieves high-rank $\mathbf{X}_q^\star$. Obtaining the optimal $q^\star \! \triangleq \! q_0 \! =\! \arg\max_{q} \text{Tr}\{\mathbf{A}_{q,1}^\prime\mathbf{X}_q^\star\} + \bar{c}_q^\prime$, problem (\ref{Prob_AsymptMaxMinEqui_rankReduct}) can be formulated. Applying the RR \cite[Algorithm 1]{HP10} to (\ref{Prob_AsymptMaxMinEqui_rankReduct}) yields rank-1 solutions. As the RR procedure is deterministic, solving (\ref{EquiProb_Approx_AsymptMaxMinSDR}) (and yielding rank-1 solutions) iteratively achieves a stationary point of (\ref{EquiProblem_AsymptMaxMin}). Due to space constraint, the algorithm is not outlined in pseudocode.

\subsubsection{Randomized CHE Max-Min Algorithm}\label{SecRandomizedCHeMaxMin}
In the above algorithm, the rank-1 solution of (\ref{Prob_AsymptMaxMinEqui_rankReduct}) is achieved by an iterative RR procedure. The following Theorem \ref{TheoProb_AsymptMaxMinEqui_RandRank1} then reveals that such a rank-1 solution can also be found by \emph{one randomized} step. Based on this, we propose a randomized CHE max-min algorithm. In the following, for notational simplicity, $\mathbf{A}_{q,1}^\prime$ (for $q \neq q_0$), $- \mathbf{A}_{q_0,1}^\prime$ and $\Lambda_q\mathbf{I}$ in (\ref{Prob_AsymptMaxMinEqui_rankReduct}) are respectively designated as $\mathbf{B}_{1,q}$, $\mathbf{B}_{1,q_0}$ and $\mathbf{B}_{2,q}$, such that (\ref{Prob_AsymptMaxMinEqui_rankReduct}) can be recast as
\begin{IEEEeqnarray}{cl}
\label{Prob_AsymptMaxMinEqui_RandRank1}
\min_{\{\!\mathbf{X}_{\!q}\! \succeq\! 0\!\}_{q\! =\! 1}^K}\,  & -\text{Tr}\{\mathbf{B}_{1,q_0} \mathbf{X}_{q_0}\} \IEEEyesnumber\IEEEyessubnumber \\
\text{s.t.}\, & \text{Tr}\!\left\{\!\mathbf{B}_{\!1,q} \mathbf{X}_{\!q}\!\right\}\! +\! \text{Tr}\!\left\{\!\mathbf{B}_{\!1,q_0} \mathbf{X}_{\!q}\!\right\}\! \leq\! \bar{c}_{\!q_0}^\prime\! -\! \bar{c}_{\!q}^\prime,\! \forall \! q\! \neq\! q_0, \IEEEyessubnumber\\
& \textstyle{\sum_{q=1}^K} \text{Tr}\{\mathbf{B}_{2,q} \mathbf{X}_q\} = 1\,.\IEEEyessubnumber
\end{IEEEeqnarray}
\begin{theorem}
\label{TheoProb_AsymptMaxMinEqui_RandRank1}
Suppose that $\{\mathbf{X}_q^\star\}_{q=1}^K$ are the high-rank optimal solutions of problem (\ref{Prob_AsymptMaxMinEqui_RandRank1}). Denote $\mathbf{X}_q^{\star1/2}$ as the Hermitian square root matrix of $\mathbf{X}_q^\star = \mathbf{X}_q^{\star1/2} \mathbf{X}_q^{\star1/2}$ and perform the EVD $\mathbf{X}_q^{\star1/2} \mathbf{B}_{1,q} \mathbf{X}_q^{\star1/2} = \mathbf{U}_q \mathbf{\Sigma}_q \mathbf{U}_q^H$. Then, there are randomized vectors $\mathbf{v}_q \! \in \! \mathbb{C}^N$ $\forall q$, such that the rank-1 matrices $\mathbf{X}_{q,\text{r1}}\! =\! \mathbf{x}_{q,\text{r1}} \mathbf{x}_{q,\text{r1}}^H$ (where $ \mathbf{x}_{q,\text{r1}}\! \triangleq \mathbf{X}_q^{\star1/2} \mathbf{U}_q \mathbf{v}_q $) are the globally optimal solutions of (\ref{Prob_AsymptMaxMinEqui_RandRank1}).
\end{theorem}
\begin{IEEEproof}
See Appendix \ref{AppTheoProb_AsymptMaxMinEqui_RandRank1} for details.
\end{IEEEproof}

\begin{algorithm}
\caption{Randomized CHE Max-Min Algorithm}\label{AlgMaxMinAsympRand}
\begin{algorithmic}[1]
\State \textbf{Initialize} Set $l = 0$; initialize $\{\mathbf{p}_q^{(0)}\}_{q=1}^K$; update $\{\mathbf{t}_q^{(0)}\}_{q = 1}^K$ by (\ref{EqConst2_EquiProbAsymptWsum}); update $\bar{c}_q^\prime$, $\mathbf{C}_{q,1}^\prime$ and $\mathbf{A}^\prime_{q,1}$ by (\ref{EqCqPrime}), (\ref{EqCq1prime}) and (\ref{EqAq1prime}), respectively; update $- \gamma_2^{\prime (0)} = \max_q \text{Tr}\{\mathbf{A}_{q,1}^\prime \mathbf{p}_q^{(0)} [\mathbf{p}_q^{(0)}]^H\} + \bar{c}_q^{\prime}$;
\Repeat
    \State $l = l + 1$;
    \State For $q \!=\! 1, \ldots, K$, update $\bar{c}_q^\prime$, $\mathbf{C}_{\!q,1}^\prime$ and $\mathbf{A}^\prime_{q,1}$ with $\mathbf{t}_q^{(l-1)}$  by (\ref{EqCqPrime}), (\ref{EqCq1prime}) and (\ref{EqAq1prime}), respectively; Solve problem (\ref{EquiProb_Approx_AsymptMaxMinSDR}) by the interior-point method, yielding high-rank $\mathbf{X}_q^\star \, \forall q$;
    \State Find the optimal $q^\star \! \triangleq q_0\! =\! \arg\max_{q} \text{Tr}\{\mathbf{A}_{q,1}^\prime\mathbf{X}_q^\star\} + \bar{c}_q^\prime$; update $\mathbf{B}_{1,q}$ and $\mathbf{B}_{2,q}$ $\forall q$ for problem (\ref{Prob_AsymptMaxMinEqui_RandRank1});
    \State Compute $\mathbf{Q}_q\! \triangleq\! \mathbf{U}_q ^H \mathbf{X}_q^{\star1/2} \mathbf{B}_{2,q} \mathbf{X}_q^{\star1/2} \mathbf{U}_q \forall q$; input $\mathbf{Q}_q$ into \cite[Algorithm 3]{HP14}, yielding \emph{randomized} $\mathbf{v}_q$; update $\mathbf{p}_q^{(l)} \! =\! \mathbf{X}_q^{\star1/2} \mathbf{U}_q \mathbf{v}_q $ according to Theorem \ref{TheoProb_AsymptMaxMinEqui_RandRank1}.
    \State Update $- \gamma_2^{\prime (l)} = \max_q \text{Tr}\{\mathbf{A}_{q,1}^\prime \mathbf{p}_q^{(l)} [\mathbf{p}_q^{(l)}]^H\} + \bar{c}_q^{\prime}$;
\Until{|- \gamma_2^{\prime (l)} - (- \gamma_2^{\prime (l-1)})|/|- \gamma_2^{\prime (l)}| \leq \epsilon} \label{AlgMaxMinAsympRand_LineStopCri}
\State $\mathbf{\bar{s}}_n^\star = \sum_{q = 1}^K [\mathbf{p}_q^{(l)}]_{n,1} \mathbf{h}_{q,n}^\ast/\sqrt{M}$ and $\mathbf{s}_\text{asym}^\star = \sqrt{E/M}\mathbf{\bar{s}}^\star/\|\mathbf{\bar{s}}^\star\|$
\end{algorithmic}
\end{algorithm}
Based on Theorem \ref{TheoProb_AsymptMaxMinEqui_RandRank1}, the randomized algorithm is proposed as Algorithm $\ref{AlgMaxMinAsympRand}$. Note that the stoping criterion at Line \ref{AlgMaxMinAsympRand_LineStopCri} of Algorithm \ref{AlgMaxMinAsympRand} is different from that in Algorithm \ref{AlgAsymptWsum} and designed based on the convergence of the objective function. This is due to the fact that the randomized $\mathbf{v}_q$ make the minimizers $\{\mathbf{p}_q^{(l)}\}_{l=1}^\infty \,\forall q$ fail to converge to limit points. Therefore, in contrast to the CHE max-min Algorithm with RR, Algorithm $\ref{AlgMaxMinAsympRand}$ may not converge to a stationary point of (\ref{EquiProblem_AsymptMaxMin}).

\section{Performance Evaluations}\label{SecSimResults}
The simulations consider a typical large open space indoor (or outdoor) WiFi-like environment at a central frequency of $2.4$\,GHz with 10\,MHz bandwidth and uncorrelated spatial domain channels, where the path loss exponent is taken from \cite{ESetal04}. The power delay profile of the IEEE TGn NLOS channel model E\cite{ESetal04} is used to generate the frequency-selective fading channel. The path loss of all the $K$ users is set as 60.046dB (for a distance of 10m with 0dB transmit/receive antenna gains), unless otherwise stated. In the following Sections \ref{SecSimResults_PwrConst} and \ref{SecSimResults_EIRPConst}, the proposed waveform designs are studied under transmit power constraints and an EIRP constraint i.e. $MP = 36$\,dBm, respectively.

\begin{table*}[!hbpt]
\centering
\caption{Shortened Names of Algorithms}
\label{Tab_Abbrev}
\begin{tabular}{|c||c|c|c|c|}
\hline
Algorithm & Algorithm \ref{AlgSCA_SU} & Algorithm \ref{AlgSCA} & Simplified weighted sum $v_\text{out}$ maximization in Section \ref{SecSimpWSum} & Algorithm \ref{AlgAsymptWsum} \\[0.3ex]
\hline
Names & SU WPT & WSum & WSum-S & CHE WSum \\ [0.3ex]
\hline \hline
Algorithm & Algorithm \ref{AlgMaxMinKleq3} & Algorithm \ref{AlgMaxMinRand_ArbiK} & CHE max-min algorithm with RR in Section \ref{SecCHeMaxMinRR} & Algorithm \ref{AlgMaxMinAsympRand} \\ [0.3ex]
\hline
Names & Max-Min-RR & Max-Min-Rand & CHE Max-Min-RR & Randomized CHE Max-Min \\ [0.3ex]
\hline
\end{tabular}
\end{table*}
Throughout Section \ref{SecSimResults}, we use shortened names as shown in Table \ref{Tab_Abbrev} (at the top of the next page) to refer to the algorithms investigated in Sections \ref{SecSU_WO}, \ref{SecWaveOptAlgos} and \ref{SecMaxMinAlgos}. These names highlight the characteristics of the algorithms. For instance, SU represents single-user; WSum and Max-Min mean that the algorithms are designed with weighted sum and max-min criteria, respectively; CHE is the abbreviation for channel hardening-exploiting; RR means that rank reduction is exploited in an algorithm to obtain a rank-1 solution. For the difference in the meaning of Rand and Randomized, please refer to Sections \ref{SecSCAthenRand} and \ref{SecRandomizedCHeMaxMin}, respectively.

\subsection{SU WPT under Transmit Power Constraints}\label{SecSimResults_PwrConst}
\begin{figure}[t]
\centering
\subfigure[]{
\label{Fig_SU_WPT_Vout_Dist}
\includegraphics[width = 2.7in]{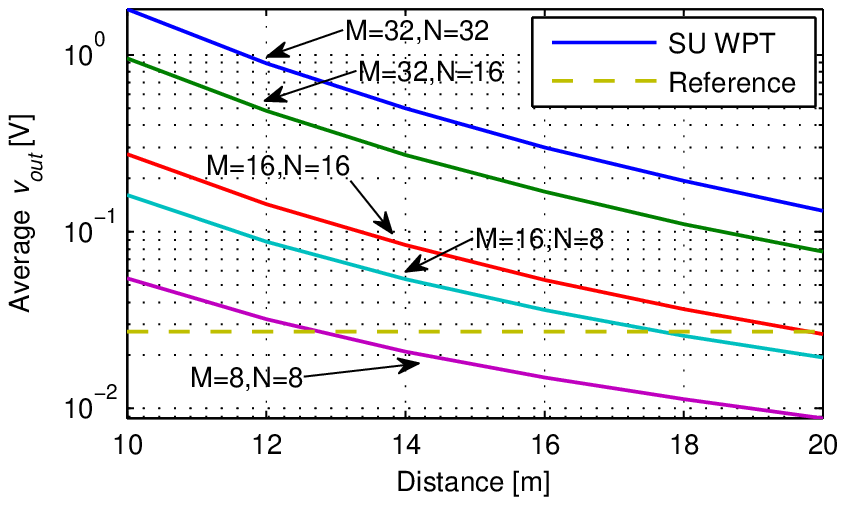}
}
\subfigure[]{
\label{Fig_ASS_Vout_Dist}
\includegraphics[width = 2.7in]{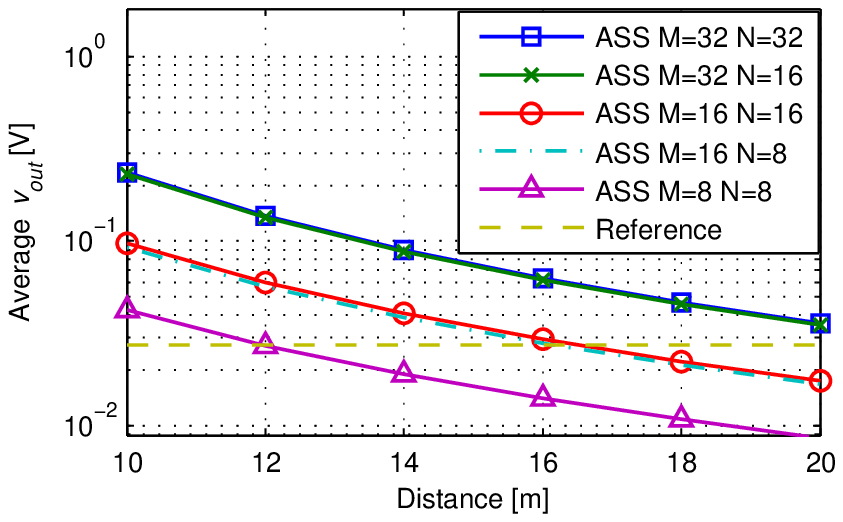}
}
\caption{(a) Average $v_\text{out}$ achieved by SU WPT as a function of distance. (b) Average $v_\text{out}$ achieved by ASS as a function of distance.}
\label{Fig_Multisine_ASS_WPT_Vout_Dist}
\end{figure}
Fig. \ref{Fig_Multisine_ASS_WPT_Vout_Dist} investigates the contribution of increasing $M$ or $N$ to the enhancement of WPT operating range, where the transmit power is subject to $P = 0.5$\,W. Figs. \ref{Fig_SU_WPT_Vout_Dist} and \ref{Fig_ASS_Vout_Dist} investigate the average $v_\text{out}$ achieved by SU WPT and Adaptive Single Sinewave (ASS) \cite{CBmar16TSParxiv} as a function of distance, respectively. ASS allocates power to the frequency with respect to the highest channel power gain. ASS is optimal for maximizing the $v_\text{out}$ formulated as a linear model (i.e. a 2nd-order truncation model\cite{Wetenkamp83, ZZH13, CBmar16TSParxiv} which is the term in (\ref{EqFuncVoutq}) with respect to $\beta_2$). The average $v_\text{out,ref} = 0.02734$\,V achieved by SU WPT for $M = 8$, $N=1$ and a distance of 10\,m is set as a reference.
Fig. \ref{Fig_SU_WPT_Vout_Dist} indicates that by increasing $M$ or $N$, SU WPT can achieve the reference $v_\text{out, ref}$ at a larger distance.
Comparing Fig. \ref{Fig_SU_WPT_Vout_Dist} to Fig. \ref{Fig_ASS_Vout_Dist} provides insights into the gain of the non-linear model-based design (i.e. SU WPT) over the linear model-based design (i.e. ASS).
It can be seen from Figs. \ref{Fig_SU_WPT_Vout_Dist} and \ref{Fig_ASS_Vout_Dist} that given $M$ and a certain distance, increasing $N$ leads to a significant increase in the average $v_\text{out}$ achieved by SU WPT, while the increase in the average $v_\text{out}$ offered by ASS is negligible.
It can be drawn from the comparison of Figs. \ref{Fig_SU_WPT_Vout_Dist} and \ref{Fig_ASS_Vout_Dist} that given $M$ and $N$, SU WPT always achieves the reference $v_\text{out,ref}$ at a larger distance than ASS, i.e. the linear model-based waveform design. For instance, given $M=16$ and $N=16$, SU WPT can achieve an average $v_\text{out} \geq v_\text{out,ref}$ at a distance $\leq 20$\,m, while ASS can achieve an average $v_\text{out} \geq v_\text{out,ref}$ at a distance $\leq 16$\,m.

Table \ref{Tab_SUWPT_vs_ReversedGP} compares the computational efficiency for SU WPT and the reversed GP-based waveform optimization \cite{CBmar16TSParxiv}.
To draw the comparison, the stopping criteria of SU WPT and the reserved GP \cite{CBmar16TSParxiv} are designed as the relative gap between the $v_\text{out}$ obtained in adjacent iterations being less than a threshold\footnote{Recall that as shown in Appendix \ref{AppTheoConvergenceAlgSCA_SU}, the objective function of problem (\ref{ApproxConvProblem_MaxVoutSU}) is monotonically decreasing, i.e. $\gamma_0^{(l)} \leq \gamma_0^{(l-1)}$. As $v_\text{out}$ is bounded, it can be shown that for SU WPT, $v_\text{out}^{(l)}$ converges as $l$ tends to infinity.}, i.e. $(v_\text{out}^{(l)} - v_\text{out}^{(l-1)})/v_\text{out}^{(l)} \leq 10^{-3}$.
Additionally, we employ the same initial point for the two algorithms.
The simulation is conducted for 100 channel realizations, accounting for $M=1$, $N=8$ and a transmit power constraint $P = 3.98107$\,W. The simulation is conducted by MATLAB R2013a on a single computer, with an Intel Core i7 processor at 3.4GHz, a RAM of 8GB and the Windows 7 Enterprise Service Pack 1 operating system.
Simulation results indicate that SU WPT converges more quickly than the reversed GP. The average elapsed time for SU WPT is significantly less than that for the reversed GP. Even though both algorithms converge to a stationary point, $v_\text{out}$ achieved by SU WPT is slightly higher than that achieved by the reversed GP, thanks to the faster convergence.

\begin{table}[t]
\centering
\caption{Elapsed running time: SU WPT vs. Reversed GP}
\label{Tab_SUWPT_vs_ReversedGP}
\begin{tabular}{|c|c|c|c|}
\hline
Algorithms & Average $v_\text{out}$ [V] &
    \begin{tabular}{@{}c@{}} {\scriptsize Average elapsed} \\  {\scriptsize time [s]} \end{tabular}
    & \begin{tabular}{@{}c@{}} {\scriptsize Average} \\  {\scriptsize convergence time} \end{tabular}\\ [0.3ex]
\hline
SU WPT & $9.532 \times 10^{-2}$ & $1.752 \times 10^{-3}$ & 4.18 iterations \\ [0.3ex]
\hline
Reversed GP & $8.417 \times 10^{-2}$ & 99.04 & 17.16 iterations \\ [0.3ex]
\hline
\end{tabular}
\end{table}

\subsection{Weighted Sum Algorithms under the EIRP Constraint}\label{SecSimResults_EIRPConst}
\begin{figure}[t]
\centering
\includegraphics[width = 2.7in]{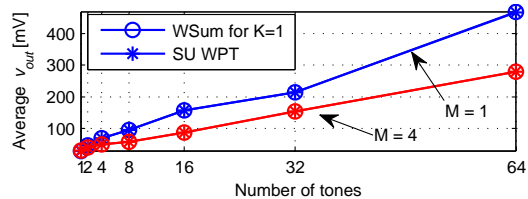}
\caption{Average $v_\text{out}$ as a function of number of sinewaves.}
\label{Fig_Comp_SU_WPT_WSumK1}
\end{figure}
Fig. \ref{Fig_Comp_SU_WPT_WSumK1} compares the average $v_\text{out}$ performance of SU WPT and WSum, in the presence of $K=1$, for an EIRP fixed to $36$\,dBm. Recall the discussion in Section \ref{SecGeneWirelessChan} that WSum optimizes both the normalized spatial domain beamforming and the power allocation across frequencies, while SU WPT only optimizes the latter. Despite this, it is shown that SU WPT achieves the same performance as WSum. This confirms Theorem \ref{TheoSUwptMRT}. As EIRP is fixed, the average $v_\text{out}$ of $M=4$ is lower than that of $M=1$.

\begin{figure}[t]
\centering
\subfigure[]{
\label{Fig_Vout_vs_MN_SU-WPT_CHE-WSum_UP_ASS_M1}
\includegraphics[width = 2.7in]
{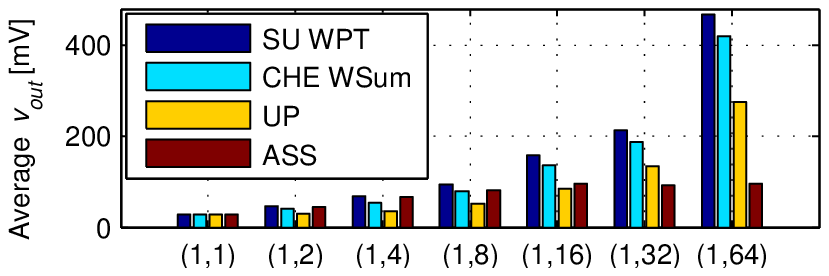}
}
\subfigure[]{
\label{Fig_Vout_vs_MN_SU-WPT_CHE-WSum_UP_ASS_M4}
\includegraphics[width = 2.7in]
{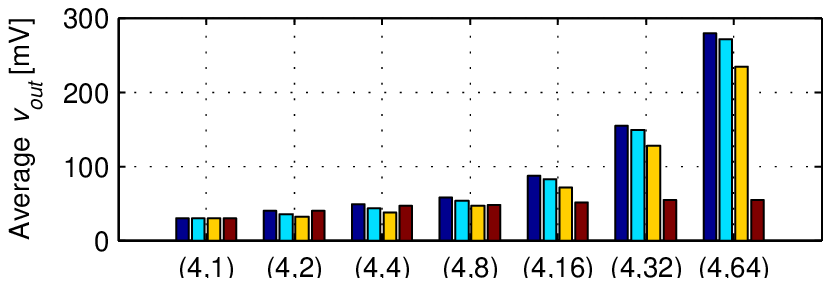}
}
\subfigure[]{
\label{Fig_Vout_vs_MN_SU-WPT_CHE-WSum_UP_ASS_M20}
\includegraphics[width = 2.7in]
{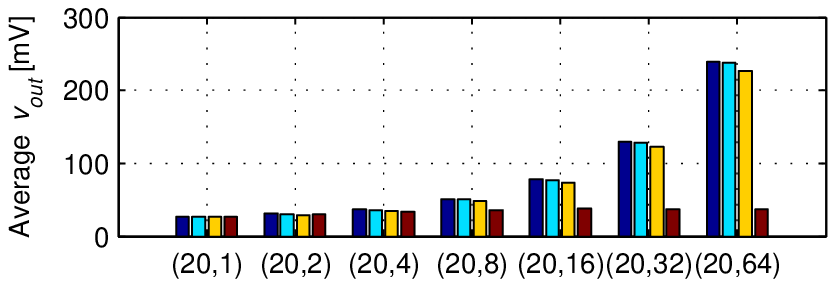}
}
\caption{Average $v_\text{out}$ as a function of $(M, N)$ with $K = 1$.}
\label{Fig_Vout_vs_MN_SU-WPT_CHE-WSum_UP_ASS}
\end{figure}
We then study the $v_\text{out}$ performance of SU WPT and CHE WSum, for $K=1$ and an EIRP fixed to $36$\,dBm, i.e. $MP = 36$\,dBm. We also consider two baseline schemes: Uniform Power (UP) allocation and ASS \cite{CBmar16TSParxiv}, both of which use MRT as the normalized spatial domain beamforming but different frequency domain power allocation. UP simply allocates power uniformly across frequencies.
Fig. \ref{Fig_Vout_vs_MN_SU-WPT_CHE-WSum_UP_ASS} shows that given $M$, the performance gain achieved by SU WPT over ASS scales with $N$ and becomes significantly large.
The observation is due to the fact that with a fixed bandwidth, as $N$ increases, adjacent frequency domain channel power gains are more likely to be distributed within a narrower range\footnote{The audience can refer to \cite[Fig. 4]{CBmar16TSParxiv} for an example of the frequency response of a 10 MHz bandwidth channel and the corresponding power allocation}. Hence, allocating all the power to the strongest frequency domain channel can be strictly suboptimal.
This observation further confirms \cite{CBmar16TSParxiv} and highlights the inaccuracy of modeling a rectifier by using a linear model. The observation also depicts that by increasing $N$, the energy transfer efficiency (i.e. $v_\text{out}/P$) achieved by the nonlinear model-based design (i.e. SU WPT) can be significantly higher than that achieved by the linear model-based design (i.e. ASS).
\begin{table}[t]
\centering
\caption{Energy transfer efficiency $\eta_t$ [V/W] for $N=16$}
\label{Tab_EnergyTransEffiency}
\begin{tabular}{|c|c|c|c|}
\hline
Algorithms & $M = 1$ & $M = 4$ & $M = 20$ \\ [0.3ex]
\hline
SU WPT & 0.0397 & 0.0873 & 0.3914 \\ [0.3ex]
\hline
ASS & 0.0242 & 0.0508 & 0.1894 \\ [0.3ex]
\hline
\end{tabular}
\end{table}
Moreover, it can be seen from Fig. \ref{Fig_Vout_vs_MN_SU-WPT_CHE-WSum_UP_ASS} that given $N$, the energy transfer efficiency (i.e. $v_\text{out}/P$) scales with the increasing $M$, though the transmit power decreases with the increasing $M$ (due to $MP = 36$\,dBm). To highlight this observation, Table \ref{Tab_EnergyTransEffiency} demonstrates the energy transfer efficiency $\eta_t \triangleq v_\text{out}/P$ values derived from the data in Fig. \ref{Fig_Vout_vs_MN_SU-WPT_CHE-WSum_UP_ASS} for $N=16$. The data in Table \ref{Tab_EnergyTransEffiency} confirms that the increasing $M$ improves $\eta_t$, and the $\eta_t$ achieved by SU WPT is significantly higher than that offered by ASS.
Fig. \ref{Fig_Vout_vs_MN_SU-WPT_CHE-WSum_UP_ASS} also shows that CHE WSum, where the iterative optimization of $\xi_{1,n}$ (which is related to the frequency domain power allocation) relies on large scale fading CSIT, can offer similar performance to SU WPT for sufficiently large $M$. However, it is worth noting that compared to SU WPT, CHE WSum can be performed less frequently due to the iterative optimization of $\xi_{1,n}$ in CHE WSum exploiting statistical CSIT based on large-scale fading, which usually varies slowly.
We also observe that given $M$, the $v_\text{out}$ performance of CHE WSum and UP can be enhanced by increasing $N$, while the performance gain of CHE WSum over UP can be enlarged by increasing $N$.
This also suggests that the scaling law of WPT\cite{CBmar16TSParxiv} offers a lower bound.

\begin{figure}[t]
\centering
\includegraphics[width = 2.6in]{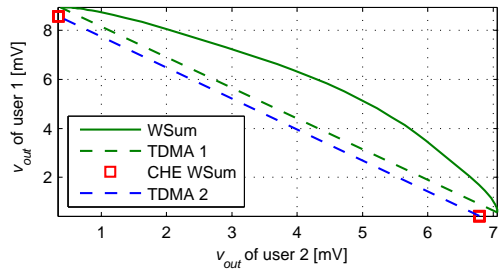}
\caption{Achievable $v_\text{out}$ region, with $M=20$ and $N=10$.}
\label{Fig_VoutRegion}
\end{figure}
We then investigate the achievable $v_\text{out}$ region of a two-user system, where the path loss of the two users is set as $66.07$\,dB (corresponding to a distance of 20m). The achievable regions of WSum and CHE WSum are obtained by performing the corresponding algorithm for a given channel realization with various user weight pairs $(w_1,w_2)$.
As a reminder, WSum optimize the waveform precoder $\mathbf{s}$, relying on small-scale fading CSI.
By contrast, the optimization of $\xi_{q,n}$ (which is related to the power allocation in the frequency domain and across the users) in CHE WSum relies on large-scale fading, although line \ref{AlgoLine_CHE_WSum_wfPrecoder} in CHE WSum (Algorithm \ref{AlgAsymptWsum}) implies that the waveform precoder at each frequency is still a function of $\mathbf{h}_{q,n}$, relying on short-term CSI.
In the legend of Fig. \ref{Fig_VoutRegion}, TDMA 1 (or TDMA 2) means that the two users are served in different timeslots; when a user is served, the waveform precoder $\mathbf{s}^\star$ (or $\mathbf{s}_\text{asym}^\star$) is optimized by WSum (or CHE WSum).
Fig. \ref{Fig_VoutRegion} demonstrates that the achievable region of WSum is larger than that of TDMA 1. This is due to the fact that WSum serves the two users simultaneously, aiming at maximizing a weighted sum criterion.
It is also shown that TDMA 2 is outperformed by TDMA 1. This comes from the fact that TDMA 2 does not exploit the small-scale fading CSI for power optimization across frequencies and users. Note that CHE WSum only obtains two $v_\text{out}$ pairs. This is explained as follows. CHE WSum is essentially a function of $\Lambda_q$ and $w_q$. For $\Lambda_1 = \Lambda_2$, the solution provided by CHE WSum only relies on $(w_1,w_2)$. When $w_1\! \neq \! w_2$, all the power is always allocated to $\mathbf{p}_q^\star$ with $q^\star = \arg \max_{q} w_q$. This is equivalent to the TDMA scenario where only one user is served. On the other hand, when $w_1\! =\! w_2$, all the power is randomly allocated to either $\mathbf{p}_1$ or $\mathbf{p}_2$. The comparison of the $v_\text{out}$ regions achieved by WSum and CHE WSum (also those achieved by TDMA 1 and TDMA 2) highlights the significance of exploiting small-scale fading CSI in the entire multi-user waveform design.

\begin{figure}[t]
\centering
\includegraphics[width = 2.6in]{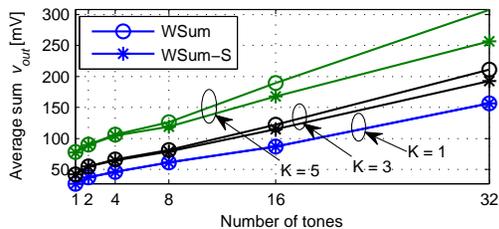}
\caption{Average sum $v_\text{out}$ as a function of $N$, with $M = 4$.}
\label{Fig_Comp_WSum_vs_WSum-S}
\end{figure}
Fig. \ref{Fig_Comp_WSum_vs_WSum-S} compares the average sum $v_\text{out}$ achieved by WSum to that achieved by WSum-S, which features a linear model-based spatial domain beamforming design but a nonlinear model-based frequency domain power optimization. In the simulations, the weight for each user is set as $w_q = 1$. Recall the discussion in Section \ref{SecSimpWSum} that WSum-S performs optimization in the frequency domain, while fixes the spatial domain beamforming $\mathbf{w}_n$ as the dominant eigenvector of $\sum_{q = 1}^K w_q \mathbf{h}_{q,n}^\ast \mathbf{h}_{q,n}^T$. Intuitively, for $K=1$, $\mathbf{w}_n$ satisfies Theorem \ref{TheoSUwptMRT}. Therefore WSum-S should be equivalent to SU WPT and offer the same performance as WSum. Motivated by this intuition, we plot WSum and WSum-S for $K=1$ in Fig. \ref{Fig_Comp_WSum_vs_WSum-S}, and the observation confirms the intuition.
Additionally, it is shown that for $K > 1$, the performance gap between WSum and WSum-S increases as $N$ increases, while the gap is negligible in the presence of a small $N$. Given $N$, the gap can also scale with $K$. The performance gap indicates that for $K>1$, a multi-user waveform strategy accounting for the linear model even only in the spatial domain beamforming design can still be significantly outperformed by the nonlinear model-based multi-user waveform design, in terms of the $v_\text{out}$ performance. Hence, the frequency domain power allocation and the spatial domain beamforming should be jointly optimized.
The observation also suggests that for small $K$ and small $N$, the less complex WSum-S can yield nearly the same $v_\text{out}$ performance as WSum.

\subsection{Max-Min Algorithms under the EIRP Constraint}
\begin{figure}[t]
\centering
\includegraphics[width = 3.0in]{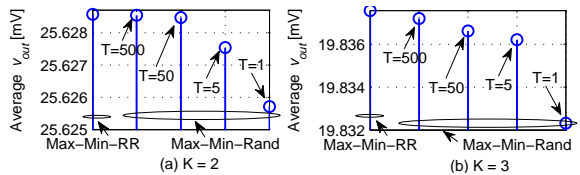}
\caption{Average minimum $v_{out}$ of Max-Min-RR and Max-Min-Rand in the presence of two users and three users, with $M=4$ and $N=4$.}
\label{Fig_Comp_MaxMimRR_MaxMinRand1_5_50_500}
\end{figure}
We first study the average minimum $v_\text{out}$ performance of Max-Min-Rand, comparing it to that of Max-Min-RR, when $K \leq 3$. Fig. \ref{Fig_Comp_MaxMimRR_MaxMinRand1_5_50_500} illustrates that Max-Min-RR always outperforms Max-Min-Rand. This is because Max-Min-RR converges to a stationary point of (\ref{Prob_MaxMinEqui_lowDeg}), as discussed in Section \ref{SecMaxMin_RR}. Despite this, the average minimum $v_\text{out}$ obtained by Max-Min-Rand is close to that of Max-Min-RR. This results from the high-rank solution $\mathbf{X}^\star$ provided by Max-Min-Rand having a dominant eigenvalue which is much larger than the other eigenvalues.
As discussed in Section \ref{SecSCAthenRand}, the solution of Max-Min-Rand is chosen from $T$ random feasible solutions. It is shown in Fig. \ref{Fig_Comp_MaxMimRR_MaxMinRand1_5_50_500} that the average minimum $v_\text{out}$ of Max-Min-Rand scales with $T$. In the following simulations, we set $T=50$.

\begin{figure}[t]
\centering
\includegraphics[width = 3.3in]{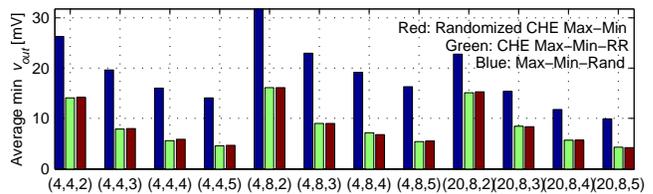}
\caption{Average minimum $v_{out}$ as a function of $(M,N,K)$.}
\label{Fig_MaxMin_Vout_vs_NumUsers_plot_TGn}
\end{figure}
Fig. \ref{Fig_MaxMin_Vout_vs_NumUsers_plot_TGn} studies the minimum $v_{out}$ performance of Randomized CHE Max-Min, CHE Max-Min-RR and Max-Min-Rand, suggesting that Randomized CHE Max-Min would be preferred for large-scale WPT. Discussion and analysis are detailed as follows.
As discussed in Section \ref{SecCHE_MaxMinAlgos}, CHE Max-Min-RR converges to a stationary point of (\ref{EquiProblem_AsymptMaxMin}), while the minimizers of Randomized CHE Max-Min cannot even converge to a limit point. Intuitively, CHE Max-Min-RR is more likely to outperform Randomized CHE Max-Min in terms of the minimum $v_\text{out}$. Nevertheless, Fig. \ref{Fig_MaxMin_Vout_vs_NumUsers_plot_TGn} reveals that CHE Max-Min-RR and Randomized CHE Max-Min offer similar performance. Due to this and the fact that Randomized CHE Max-Min computes the rank-1 solution of (\ref{Prob_AsymptMaxMinEqui_rankReduct}) by one randomized step, Randomized CHE Max-Min is more computationally efficient than CHE Max-Min-RR.
Furthermore, the comparison of the average minimum $v_\text{out}$ with $(M, N) = (4, 4)$ and that with $(M, N) = (4, 8)$ shows that increasing $N$ benefits the max-min algorithms, in terms of the average minimum $v_\text{out}$.
Fig. \ref{Fig_MaxMin_Vout_vs_NumUsers_plot_TGn} shows that given $(N, K)$, the performance gap between Max-Min-Rand and the CHE max-min algorithms (including CHE Max-Min-RR and Randomized CHE Max-Min) decreases, as $M$ increases. This implies that the CHE Max-Min algorithms and Max-Min-Rand can offer similar performance, when $M$ is sufficiently large. Due to this and the lower complexity, Randomized CHE Max-Min would be preferred for large-scale designs.

\begin{figure}[t]
\centering
\includegraphics[width = 2.8in]{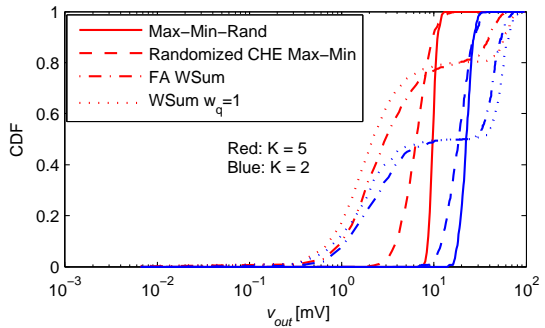}
\caption{CDF of $v_\text{out}$, with $M=20$ and $N=8$.}
\label{Fig_CDF_TGn}
\end{figure}
We now study the CDF performance of Max-Min-Rand and Randomized CHE Max-Min, considering WSum with $w_q = 1\, \forall q$ as a baseline scheme. We also consider a Fairness-Aware (FA) WSum algorithm as another baseline. Intuitively, the larger weights in FA WSum should be assigned to the users suffering from lower channel power gains. Due to the fact that $v_\text{out}$ is also a nonlinear function of spatial/frequency domain channel gains (for given transmit waveforms), an indicator is designed to jointly evaluate user $q$'s spatial and frequency domain channels. Specifically, assume that only user $q$ is served, the output voltage achieved by UP (i.e. the baseline in Fig. \ref{Fig_Vout_vs_MN_SU-WPT_CHE-WSum_UP_ASS}) is utilized as the indicator for user $q$ and designated as $\alpha_q$. Therefore, the fairness-aware weight $w_q = \alpha_q^{-1}/(\sum_{q=1}^K \alpha_q^{-1})$. Fig. \ref{Fig_CDF_TGn} demonstrates that Max-Min-Rand outperforms other algorithms in terms of fairness. Numerical results also confirm that FA WSum offers fairer $v_\text{out}$ than WSum. It is worth noting that although the optimization in Randomized CHE Max-Min could not leverage CSIT on small-scale fading, the algorithm still demonstrates better fairness than FA WSum.

\begin{figure}[t]
\centering
\includegraphics[width = 2.7in]{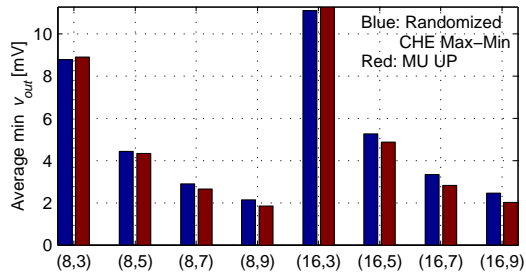}
\caption{Average minimum $v_\text{out}$ as a function of $(N,K)$, with $M = 50$.}
\label{Fig_AvgMinVout_vs_N8_16K}
\end{figure}
\begin{figure}[t]
\centering
\includegraphics[width = 2.7in]{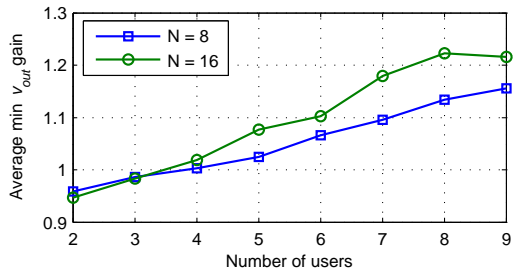}
\caption{Average minimum $v_\text{out}$ gain of Randomized CHE Max-Min over MU UP as a function of $K$, with $M = 50$.}
\label{Fig_AvgMinVoutGain}
\end{figure}
We then investigate the average minimum $v_\text{out}$ performance of Randomized CHE Max-Min, considering larger $M$ and $N$. The baseline is a Multi-User (MU) Uniform Power (UP) allocation scheme, which exploits the asymptotically optimal spatial domain beamforming and uniformly allocates the transmit power across frequencies and users, such that the precoder at frequency $n$ is $\mathbf{s}_n = \sqrt{P} \mathbf{w}_n/\sqrt{\sum \|\mathbf{w}_n\|^2}$ for $\mathbf{w}_n = \sum_{q = 1}^K \mathbf{h}_{q,n}^\ast/\|\mathbf{h}_{q,n}\|$. By contrast, Randomized CHE Max-Min not only allocates more power to the vector\footnote{The vector $\mathbf{p}_q$ is an optimization variable of problem (\ref{EquiProblem_AsymptMaxMin}), defined below (\ref{EqAsymptOutputVol}) and related to the fraction of transmit power allocated for user $q$.} $\mathbf{p}_q$ corresponding to a lower $\Lambda_q$, but also optimizes power allocation across frequencies, such that all the users achieve the same asymptotical output voltage.
Note that in the simulations of Figs. \ref{Fig_AvgMinVout_vs_N8_16K} and \ref{Fig_AvgMinVoutGain}, Randomized CHE Max-Min uniformly allocates power across users, due to the same $\Lambda_q\, \forall q$.
Fig. \ref{Fig_AvgMinVout_vs_N8_16K} shows that when $K$ is small, as channel fluctuations contribute to $v_\text{out}$, Randomized CHE Max-Min can be outperformed by MU UP. As $K$ increases, Randomized CHE Max-Min can outperform MU UP, in terms of the average minimum $v_\text{out}$. This performance gain of Randomized CHE Max-Min over MU UP results from optimizing power allocation across frequencies, since Randomized CHE Max-Min uniformly allocate power across users (due to the same $\Lambda_q\, \forall q$) as MU UP does.
Moreover, given $K$, comparing the average minimum $v_\text{out}$ performance for $N=8$ and $N=16$ reveals that both Randomized CHE Max-Min and MU UP can benefit from the increasing $N$, yielding higher average minimum $v_\text{out}$.
Fig. \ref{Fig_AvgMinVoutGain} presents the average minimum $v_\text{out}$ gain of Randomized CHE Max-Min over MU UP, in the presence of $N = 8$ and $N = 16$. It is shown that the curve of $N = 16$ is higher than that of $N = 8$ when $K$ is sufficiently large, and the gap between the two curves increases as $K$ increases.
The observation indicates that the minimum $v_\text{out}$ performance gain of frequency domain power optimization over uniformly allocating power across frequencies can be enlarged by increasing $N$, and the improvement on this gain is more significant in the presence of a large $K$.

\section{Conclusions}\label{SecConclu}
This paper studies waveform optimizations for the large-scale multi-antenna multi-sine Wireless Power Transfer (WPT). We have developed an optimization framework towards low-complexity algorithms for the problems involving the nonlinear rectenna model. Single-user and multi-user algorithms are designed to maximize the weighted-sum/minimum rectenna DC output voltage. The study highlights the effect of the nonlinearity on the waveform design. It is revealed that the single-user WPT design boils down to optimizing power allocation across frequencies. However, in general, to maximize the weighted-sum criterion, the optimal spatial domain beamforming and the optimal frequency domain power allocation cannot be designed separately. In the presence of a larger number of antennas and sinewaves, channel hardening-exploiting algorithms are designed as well. Asymptotic analysis reveals that the structure of the asymptotical optimal spatial domain precoder can be designed prior to the power allocation optimization across users and frequencies. Accounting for the nonlinear model, the proposed multi-user algorithms can benefit from the increasing number of sinewaves, in terms of the output voltage performance.

\appendix
\subsection{Proof of Theorem \ref{TheoSUwptMRT}}\label{AppTheoSUwptMRT}
\setcounter{equation}{0}
\renewcommand{\theequation}{A.\arabic{equation}}
To prove Theorem \ref{TheoSUwptMRT}, we show that given $\xi_n$ $\forall n$, $\mathbf{\tilde{s}}_n$ has to be MRT, such that (\ref{Eqv_outInhomo}) can be maximized. It can be easily seen that to maximize $f_{\text{LPF}} \left(y_1^2(t)\right)$ i.e. $\sum_{n = 1}^N \mathbf{s}_n^H \mathbf{h}_{1,n}^\ast \mathbf{h}_{1,n}^T \mathbf{s}_n$, the optimal $\mathbf{\tilde{s}}_n = e^{j\phi_n}  \mathbf{h}_{1,n}^\ast/\|\mathbf{h}_{1,n}\|$; otherwise another $\mathbf{\tilde{s}}_n$ can always be found, yielding a higher value. Then, we will show that this $\mathbf{\tilde{s}}_n$ also maximizes $f_{\text{LPF}} \left(y_1^4(t)\right)$.
Given $n_1, n_2, n_3$ and $n_4$, as the value of $\mathbf{s}_{n_3}^H \mathbf{h}_{1,n_3}^\ast \mathbf{h}_{1,n_1}^T \cdot \mathbf{s}_{n_1} \mathbf{s}_{n_4}^H \mathbf{h}_{1,n_4}^\ast \mathbf{h}_{1,n_2}^T  \mathbf{s}_{n_2}$ is equal to that of $\big(\mathbf{s}_{n_2}^H \mathbf{h}_{1,n_2}^\ast \mathbf{h}_{1,n_4}^T \mathbf{s}_{n_4} \cdot \mathbf{s}_{n_1}^H \mathbf{h}_{1,n_1}^\ast \mathbf{h}_{1,n_3}^T  \mathbf{s}_{n_3}\big)^\ast$, (\ref{EqFlpf_yq4_2}) can be written as
\begin{equation}
\label{EqFlpf_yq4_Re}
f_{\text{LPF}}\! \left( y_1^4(t) \right) \!=\!\frac{3}{2}\! \sum\! \text{Re}\!\left\{\! \mathbf{s}_{n_3}^H \! \mathbf{h}_{1,n_3}^\ast \! \mathbf{h}_{1,n_1}^T \mathbf{s}_{n_1} \! \mathbf{s}_{n_4}^H \! \mathbf{h}_{1,n_4}^\ast \! \mathbf{h}_{1,n_2}^T \! \mathbf{s}_{n_2}\!\right\}\!,
\end{equation}
where $n_1, n_2, n_3, n_4\! \in\! \{1, \dots, N\}$ and $n_1\!  + \!  n_2\! =\! n_3\! +\! n_4$.
Note that although the value of $\mathbf{s}_{n_3}^H \mathbf{h}_{1,n_3}^\ast \mathbf{h}_{1,n_1}^T \mathbf{s}_{n_1} \mathbf{s}_{n_4}^H \mathbf{h}_{1,n_4}^\ast \mathbf{h}_{1,n_2}^T \mathbf{s}_{n_2}$ is generally complex, this value has to be real in order to maximize $f_{\text{LPF}}\! \left(y_1^4(t)\right)$, due to $\text{Re}\{\cdot\}$ in (\ref{EqFlpf_yq4_Re}).
In the following, we first show the optimality of $\mathbf{\tilde{s}}_{n_3}$ and $\mathbf{\tilde{s}}_{n_2}$, with given $\mathbf{s}_{n_1}$ and $\mathbf{s}_{n_4}$. Specifically, defining $\mathbf{\tilde{s}}_n = e^{j\phi_n} \mathbf{\tilde{w}}_n $, we shall show the optimal $\mathbf{\tilde{w}}_n = \mathbf{h}_{1,n}^\ast/\|\mathbf{h}_{1,n}\|$ for $n\! \in\! \{n_2, n_3\}$.
Assuming $\mathbf{h}_{1,n_1}^T \mathbf{s}_{n_1} [\mathbf{s}_{n_4}]^H \mathbf{h}_{1,n_4}^\ast = c_{n_1, n_4}\! \in \! \mathbb{C}$, with the structure of $\mathbf{s}_n^\star$ shown in Theorem \ref{TheoSUwptMRT}, $[\mathbf{s}_{n_3}]^H \mathbf{h}_{1,n_3}^\ast \mathbf{h}_{1,n_1}^T \mathbf{s}_{n_1} [\mathbf{s}_{n_4}]^H \mathbf{h}_{1,n_4}^\ast \mathbf{h}_{1,n_2}^T \mathbf{s}_{n_2}$ can be written as $\xi_{n_3}^\ast e^{-j\phi_{n_3}} \mathbf{\tilde{w}}_{n_3}^H \mathbf{h}_{1,n_3}^\ast c_{n_1, n_4}\mathbf{h}_{1,n_2}^T \xi_{n_2} e^{j\phi_{n_2}} \mathbf{\tilde{w}}_{n_2}$.
To maximize $\text{Re}\{\xi_{n_3}^\ast e^{j(\phi_{n_2}\!-\phi_{n_3})} \mathbf{\tilde{w}}_{n_3}^H \mathbf{h}_{1,n_3}^\ast \mathbf{h}_{1,n_2}^T \mathbf{\tilde{w}}_{n_2} \xi_{n_2} c_{n_1, n_4}\}$, the optimal $\mathbf{\tilde{w}}_{n_3}$ and $\mathbf{\tilde{w}}_{n_2}$ turn out to be $\mathbf{h}_{n_3}^\ast/\|\mathbf{h}_{n_3}\|$ and $\mathbf{h}_{n_2}^\ast/\|\mathbf{h}_{n_2}\|$, which are the left and the right singular vectors of the rank-1 matrix $\mathbf{h}_{1,n_3}^\ast\mathbf{h}_{1,n_2}^T$. Meanwhile, the value of $\phi_{n_2}\! - \phi_{n_3}$ should be determined such that $\xi_{n_3}^\ast \xi_{n_2} c_{n_1, n_4} e^{j(\phi_{n_2} - \phi_{n_3})}$ is real.
Similarly, given $\mathbf{s}_{n_2}$ and $\mathbf{s}_{n_3}$, $\mathbf{s}_{n_3}^H \mathbf{h}_{1,n_3}^\ast \mathbf{h}_{1,n_1}^T \mathbf{s}_{n_1} \mathbf{s}_{n_4}^H \cdot  \mathbf{h}_{1,n_4}^\ast \mathbf{h}_{1,n_2}^T \mathbf{s}_{n_2}$
becomes $\xi_{n_4}^\ast e^{j(\phi_{n_1} - \phi_{n_4})} \mathbf{\tilde{w}}_{n_4}^H \mathbf{h}_{1,n_4}^\ast c_{n_2, n_3} \cdot \mathbf{h}_{1,n_1}^T  \xi_{n_1} \mathbf{\tilde{w}}_{n_1}$. It can be shown that the optimal ${\mathbf{\tilde{w}}}_{n_1}$ and ${\mathbf{\tilde{w}}}_{n_4}$ are $\mathbf{h}_{n_1}^\ast/\|\mathbf{h}_{n_1}\|$ and  $\mathbf{h}_{n_4}^\ast/\|\mathbf{h}_{n_4}\|$, respectively.

\subsection{Proof of Proposition \ref{PropRank1soluSU}}\label{AppPropRank1soluSU}
\setcounter{equation}{0}
\renewcommand{\theequation}{B.\arabic{equation}}
In (\ref{Eq_C1doublePrime}), as $\mathbf{M}_0^{\prime\prime} \succ 0$ and $\mathbf{X} \succeq 0$, it follows that $t^{(l - 1)}_0 \geq 0$. Hence, $\text{Tr}\{\mathbf{A}_1^{\prime\prime}\} < 0$. This means that $\mathbf{A}_1^{\prime\prime}$ always exists an eigenvalue less than zero.
Given that problem (\ref{ApproxConvProblem_MaxVoutSU_Equiv}) yields a rank-1 solution $\mathbf{X}^\star = \mathbf{x}^\star[\mathbf{x}^\star]^H$, (\ref{ApproxConvProblem_MaxVoutSU_Equiv}) is equivalent to a nonconvex quadratically constrained quadratic problem (QCQP) given by
$\min_{\mathbf{x}} \{ \mathbf{x}^H\mathbf{A}_1^{\prime\prime}\mathbf{x}: \|\mathbf{x}\|^2 \leq P \}$.
The KKT conditions of
this problem indicates that the stationary points are in the directions of the eigenvectors of $\mathbf{A}_1^{\prime\prime}$.
Hence, $\mathbf{x}^\star = \sqrt{P}[\mathbf{U}_{\mathbf{A}_1^{\prime\prime}}]_{\text{min}}$.

\subsection{Proof of Theorem \ref{TheoConvergenceAlgSCA_SU}}\label{AppTheoConvergenceAlgSCA_SU}
\setcounter{equation}{0}
\renewcommand{\theequation}{C.\arabic{equation}}
Since $-g(\mathbf{t})$ is convex, $g(\mathbf{t}) \leq \tilde{g}(\mathbf{t}, \mathbf{t}^{(l - 1)})$. Therefore, $\tilde{g}(\mathbf{t}^{(l)}, \mathbf{t}^{(l)}) = g(\mathbf{t}^{(l)}) \leq \tilde{g}(\mathbf{t}^{(l)}, \mathbf{t}^{(l-1)})$, which indicates that the optimal solution $\mathbf{X}^{(l-1)}$ of the $(l\!-\!1)$\,th AP (\ref{ApproxConvProblem_MaxVoutSU}) is a feasible point of the $l$\,th AP (\ref{ApproxConvProblem_MaxVoutSU}). As (\ref{ApproxConvProblem_MaxVoutSU}) is convex, the objective functions over iterations satisfy $\gamma_0^{(l)} \leq \gamma_0^{(l-1)}$. Then, assuming the eigenvalue decomposition of $\mathbf{A}_1^{\prime\prime}$ is unique, we can prove that the sequence of the minimizer $\{\gamma_0^{(l)}, \mathbf{t}^{(l)}, \mathbf{X}^{(l)}\}_{l=0}^{\infty}$ converges to a limit point, by contradiction\cite[Proof of Proposition 2.7.1]{Bertsekas99}. As $\nabla g(\mathbf{t}^{(l)}) = \nabla \tilde{g}(\mathbf{t}^{(l)}, \mathbf{t}^{(l)})$, the solution of (\ref{ApproxConvProblem_MaxVoutSU}) finally converges to a stationary point of (\ref{Epi_Problem_MaxVoutSU_RankConstRelaxed}). As such a stationary point (with a rank-1 $\mathbf{X}^\star$) is also a stationary point of (\ref{EquiProb_MaxVoutSU}), Algorithm \ref{AlgSCA_SU} converges to a stationary point of (\ref{EquiProb_MaxVoutSU}).

\subsection{Proof of Theorem \ref{TheoProb_AsymptMaxMinEqui_RandRank1}}\label{AppTheoProb_AsymptMaxMinEqui_RandRank1}
\setcounter{equation}{0}
\renewcommand{\theequation}{D.\arabic{equation}}
In problem (\ref{Prob_AsymptMaxMinEqui_RandRank1}), given a certain $q$, the optimization variable $\mathbf{X}_q$ is multiplied by $\mathbf{B}_{1,q}$ and $\mathbf{B}_{2,q}$, respectively. This is reminiscent of the application of \cite[Lemma 3.3]{HP14} to the QCQP with two double-sided constraints. Here we prove Theorem \ref{TheoProb_AsymptMaxMinEqui_RandRank1} by \cite[Lemma 3.3]{HP14}: there always exists a randomized $\mathbf{v}_q$ $\forall q$, such that
{\small$\text{Tr}\{\mathbf{B}_{1,q} \mathbf{X}_{q,\text{r1}}\}\! =\! \text{Tr}\{\mathbf{B}_{1,q} \mathbf{X}_q^{\star1/2} \mathbf{U}_q \mathbf{v}_q \mathbf{v}_q^H \mathbf{U}_q ^H \mathbf{X}_q^{\star1/2} \} =$ $\text{Tr}\{\mathbf{\Sigma}_{q}$ $\mathbf{v}_q \mathbf{v}_q^H\}\! =\! \text{Tr}\{\mathbf{\Sigma}_{q} \mathbf{I}\} \!=\! \text{Tr}\{\mathbf{B}_{1,q} \mathbf{X}_q^\star\}$}. Meanwhile, {\small$\text{Tr}\{\mathbf{B}_{2,q}$ $\mathbf{X}_{q,\text{r1}}\} \! = \! \text{Tr}\{\mathbf{B}_{2,q} \mathbf{X}_q^{\star1/2} \mathbf{U}_q \mathbf{v}_q \mathbf{v}_q^H \mathbf{U}_q ^H \mathbf{X}_q^{\star1/2} \}\! = \! \text{Tr}\{\mathbf{U}_q ^H \mathbf{X}_q^{\star1/2} \mathbf{B}_{2,q} \mathbf{X}_q^{\star1/2}  \mathbf{U}_q \mathbf{v}_q  \mathbf{v}_q^H \} = \text{Tr}\{\mathbf{U}_q ^H \mathbf{X}_q^{\star1/2} \mathbf{B}_{2,q} \cdot  \mathbf{X}_q^{\star1/2} \mathbf{U}_q \mathbf{I}\} = \text{Tr}\{\mathbf{B}_{2,q} \mathbf{X}_q^\star\}$}.

\bibliographystyle{IEEEtran}
\bibliography{IEEEabrv,BibPro}

\end{document}